\documentclass[%
 reprint,
nofootinbib,
 amsmath,amssymb,
 aps,
]{revtex4-2}

\usepackage{graphicx}
\usepackage{dcolumn}
\usepackage{bm}
\usepackage{multirow}
\usepackage{amssymb,mathrsfs,amsmath}
\usepackage{amsfonts}
\usepackage{color,xcolor}
\usepackage{subfigure}
\def\linkcolor{cyan!70!black}

\usepackage[
colorlinks=true
,urlcolor=\linkcolor
,anchorcolor=\linkcolor
,citecolor=\linkcolor
,filecolor=\linkcolor
,linkcolor=\linkcolor
,menucolor=\linkcolor
,linktocpage=true
,pdfproducer=medialab
,pdfa=true
]{hyperref}

\newcommand{\bc}{\begin{center}}
\newcommand{\ec}{\end{center}}
\newcommand{\be}{\begin{equation}}
\newcommand{\ee}{\end{equation}}
\newcommand{\ba}{\begin{eqnarray}}
\newcommand{\ea}{\end{eqnarray}}
\newcommand{\bea}{\begin{eqnarray*}}
\newcommand{\eea}{\end{eqnarray*}}

\begin{document}

\title{ Effect of $K^*$ meson magnetic dipole moment on the $e^+e^- \to K^+ K^-\pi^0 \pi^0 $ cross section}

\author{Luis A. Jim\'enez P\'erez}
\affiliation{Instituto de F\'{\i}sica,  Universidad Nacional Aut\'onoma de M\'exico, AP 20-364, Ciudad de M\'exico 01000, M\'exico}%
\author{Antonio Rojas}
\affiliation{Instituto de F\'{\i}sica,  Universidad Nacional Aut\'onoma de M\'exico, AP 20-364, Ciudad de M\'exico 01000, M\'exico}%
\author{Genaro Toledo}
\affiliation{Instituto de F\'{\i}sica,  Universidad Nacional Aut\'onoma de M\'exico, AP 20-364, Ciudad de M\'exico 01000, M\'exico}%
\date{\today}

\begin{abstract}
We explore the sensitivity of the $e^{+} e^{-} \to K^+ K^- 2 \pi^0$ cross section to the magnetic dipole moment (MDM) of the $K^*$ vector meson. We describe the $\gamma^* \to 2K2\pi$ vertex using a vector meson dominance model, including the intermediate resonant contributions relevant for energies below 2.4 GeV. Using BaBar data for this process, we show that this observable is indeed sensitive to the MDM of the $K^*$; we obtain a central value for the MDM of $\mu_{K^*}=4.5$ and an upper bound of $\bar{\mu}_{K^*} =  6.3$, in units of $e/2 m_{K^*}$. We emphasize the need for higher precision data to provide a first data-driven determination of this parameter to confront it with theoretical predictions.
 
\end{abstract}

\maketitle
\section{Introduction}
The dynamics of quarks within hadrons can be explored using global observables such as electromagnetic multipoles. The proton itself was recognized as a non-fundamental through the comparison of its measured magnetic dipole moment (MDM) with the value predicted by Dirac's theory for fundamental fermions.
The MDM is usually represented by the gyromagnetic ratio, $g$, expressed in units of the particle magneton. The value $g=2$ seems to characterize not only one-half spin particles but also fundamental particles with higher spin \cite{Deser:1970spa,Brodsky:1992px,Ferrara:1992yc, Jackiw:1997vw,ToledoSanchez:2002rn, Holstein:2006wi,Dbeyssi:2011ep,Lorce:2009bs,Delgado-Acosta:2012dxv,Haberzettl:2019qpa}. 

For composite states, corrections arising from the internal dynamics are expected. Although this property has been measured for several baryons, the meson sector remains experimentally challenging, mainly due to the short lifetime of the states. A particle with spin-$s$ has $2s+1$ electromagnetic multipole moments. For mesons, vector states are therefore the simplest systems exhibiting multipoles beyond the electric charge. Namely, the magnetic dipole moment and the electric quadrupole moment. 
The $\rho$ and $K^*$ mesons are examples of such states, whose MDMs have been predicted using different approximations to QCD. Theoretical models place the value of the $\rho$ meson MDM in the range of $\mu_\rho =[1.5 \, , \, 2.7] $ in units of $e/2 m_\rho$
\cite{Bagdasaryan:1984kz,Cardarelli:1994yq,deMelo:1997hh,Melikhov:2001pm,Alexander_2003,Aliev:2003ba,Jaus:2002sv,Choi:2004ww,He:2004ba,ALIEV2009470,Biernat:2014dea,Simonis2016,deMelo:2016ynt,Krutov:2018mbu,Simonis:2018rld,DeMelo:2018bim,Hawes:1998bz, Bhagwat:2006pu,Roberts:2011wy, Pitschmann:2012by, Xu:2019ilh, Xing:2021dwe,Shi:2023oll,Xu:2023vlt, 
Haurysh2021, Zhang:2022zim, Luan_2015, Djukanovic:2013mka}.
In a recent work \cite{Rojas:2024tmn}, a value for the MDM of the $\rho$ meson was obtained ($\mu_{\rho} =  2.7 \pm 0.3 $ in units of $e/2 m_{\rho}$ ) considering experimental data for the $e^{+} e^{-} \to \pi^+ \pi^- 2 \pi^0$ process, measured by the BaBar experiment \cite{BaBar:2017zmc}. This provided the first data-based determination of this quantity, allowing  a direct comparison with theoretical predictions. This achievement was possible due to the important role played by the $\rho$ electromagnetic vertex, which contains information about the multipole structure of the radiating particle, in a specific energy region of the cross section measured by the BaBar experiment. This procedure is analogous to the one used to extract the MDM of the $W$ gauge boson  \cite{DELPHICollab,MELE2004255} from the process $e^+ e^- \to 2jets \,l \, \nu$, where $l$ and $\nu$ denote a lepton and a neutrino, respectively. 

The $K^*$ meson shares several features with the $\rho$ meson but differs by the strangeness content, which may lead to non-negligible isospin symmetry breaking effects. The predictions for the $K^*$ MDM, based on different approaches to QCD, are summarized in Table \ref{K*MDM-Table}. They are in the range of $\mu_{K^*} =[2 \, , \, 2.68] $ in units of $e/2 m_{K^*}$. However, currently no experimental information is available to compare with these predictions.

\begin{table}
	\centering
	\begin{tabular}{|c|c|c|}
		\hline MDM $(e/2M_{K^*})$& Ref. & Description \\
		\hline
		
		\hline $2.09$ & \cite{Badalian:2012ft} & Relativistic Hamiltonian \\ & &  derived from the path\\ & & integral form of the $q_1 \bar q_2$\\ &&  Green's function\\
		\hline $2.68(1)$ & \cite{Lee:2008qf} & Lattice QCD\\
		\hline$2.04\pm 0.4 $ & \cite{ALIEV2009470} & QCD sum rules \\
		\hline$2.23$ & \cite{Bhagwat:2006pu} & Poincar\'e covariant\\ && formulation based on\\ && Dyson-Schwinger \\ && equation of QCD\\
		\hline$2.37 $ & \cite{Hawes:1998bz} & Lorentz-covariant\\
        && formulation based on\\ && Dyson-Schwinger \\ && equation of QCD\\
        \hline $2-0.0047 $ & \cite{GarciaGudino:2010sd} & Finite width modification \\
		\hline
	\end{tabular}
	\caption{Predictions of the $K^{*\pm}$ meson MDM.}
	\label{K*MDM-Table}
\end{table}

The process analogous to the one used in the $\rho$ meson case is $e^{+} e^{-} \to K^+ K^- 2 \pi^0$. The cross section of this process has been measured by the BaBar Collaboration \cite{BaBar:2011btv}, where the relevant states in the low energy region, such as the $f_0(980)$ and the $K^*$ mesons, were identified trough the reconstruction of the corresponding invariant mass distribution of their decay products. To explore the sensitivity of this process to the $K^*$ MDM, it is necessary to study the role of the channel in which two intermediate $K^*$ mesons decay into $K^+ K^- 2 \pi^0$, relative to the other channels dominated by scalar resonances. 

In this work, we study the effect of the $K^*$ MDM on the $e^{+} e^{-} \to K^+ K^- 2 \pi^0$ cross section, following the same approach used in the analysis of the $\rho$ meson \cite{Rojas:2024tmn}, profiting from the BaBar analysis of the intermediate states involved. In Section II, we establish the theoretical framework of the analysis and present the multipolar structure of the $K^*$ radiative vertex. We then describe the $e^{+} e^{-} \to K^+ K^- 2 \pi^0$ process, modeling the $\gamma^* \to 2K2\pi$ vertex within the VMD approach \cite{Kroll:1967it,Bando:1984ej, Fujiwara:1984mp, Meissner:1987ge}. 
 There, we derive the electromagnetic gauge invariant amplitude of the process, decomposed into the different contributing channels, while enforcing Bose-Einstein symmetry and charge conjugation (C) invariance. 
We show that our result generalizes the isospin symmetric description obtained for the $\rho$ \cite{Rojas:2024tmn}, by introducing the isospin symmetry breaking contribution, associated with the mass difference between the kaon and the pion.
In Section III, we compute the cross section and analyze the role of the $K^*$ MDM. The parameters of the model are determined using the experimental information provided by the Particle Data Group (PDG) \cite{pdg}. Parameters with limited experimental precision, related to the scalar mesons $f_0(980)$ and $\sigma$, as well as the $K^*$ MDM are treated as free parameters and determined from a fit to BaBaR experimental data. 
In Section IV, we explore the sensitivity of the cross section to the $K^*$ MDM, using BaBar data. Due to the limited precision of the available data, only an upper bound is obtained. We conclude with a discussion of the results.

\section{Description of the $e^{+} e^{-} \to K^+ K^- 2 \pi^0$  process}
 
\begin{figure}
\begin{center}
\includegraphics[scale=0.4]{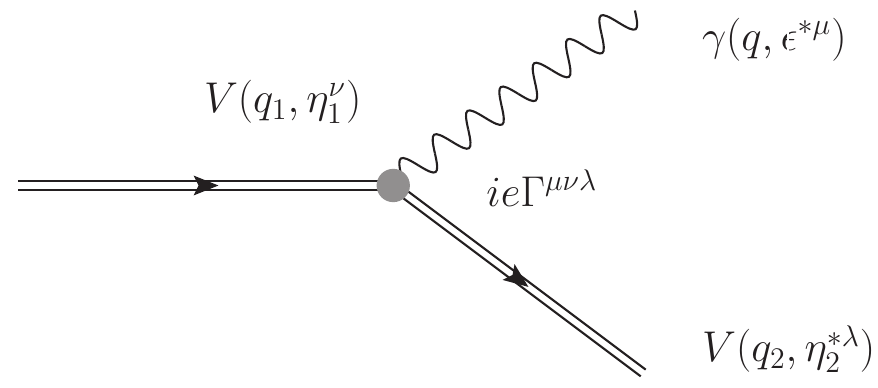}
\end{center}
\caption{$V(q_1,\eta_1)\to V(q_2,\eta_2)\gamma(q,\epsilon)$ process. The $VV\gamma$ electromagnetic vertex and momenta and polarizations convention are shown.}
\label{VVgammavertex}
\end{figure}

The radiative process of a vector particle ($V$) is illustrated in Fig. \ref{VVgammavertex}. We set the notation by $V(q_1, \eta_1) \to V(q_2, \eta_2) \gamma (q,\epsilon)$, where $\gamma$ is the photon field and in parentheses are the corresponding momentum and polarization vector. The electromagnetic vertex is defined through the matrix element
\begin{equation}
    \langle V(q_2, \eta_2)|J^\mu_{EM}(0)|V(q_1, \eta_1) \rangle \equiv e\Gamma^{\mu \nu \lambda} \eta_{1\nu} \eta^*_{2\lambda}.
\end{equation}

The $C$, $P$ (parity) and $CP$ conserving electromagnetic vertex $\Gamma^{\mu \nu \lambda}$ can be decomposed into multipolar structures as follows
\begin{eqnarray}
\Gamma^{\mu\nu\lambda}&=&
 \alpha(q^{2}) g^{\nu \lambda}(q_1 + q_2)^{\mu} + \beta(q^{2}) ( g^{\mu \nu} q^{\lambda} -  g^{\mu \lambda} q^{\nu})\nonumber\\
  &&- \frac{\gamma(q^{2})}{M_V^2}(q_1 + q_2)^{\mu} q^{\nu} q^{\lambda}- q_1^\lambda g^{\mu \nu} -q_2^\nu g^{\mu \lambda},
\label{vertex}
\end{eqnarray}
where $\alpha(q^{2})$, $\beta(q^{2})$  and $\gamma(q^{2})$ correspond to the electric charge (in units of $e$ ), magnetic dipole moment (in units of $e/2 M_V$ ), and the electric quadrupole moment (in units of $e/M_V^{2}$ ) form factors \cite{Kim:1973ee,Hagiwara:1986vm,Nieves:1996ff,Gounaris:1996rz}. In the static limit, these quantities correspond to electromagnetic multipoles.
$\alpha(0) = 1$, $\beta(0) = 2$, and  $\gamma(0) =0 $ are usually taken as reference values for vector mesons. For the $K^*$ meson, the physical values are expected to be affected by the strong interaction dynamics among quarks. Notice that the magnetic term grows as ${\cal O}(q)$ whereas the quadrupole grows as ${\cal O}(q^2)$. Thus, for a relatively small photon energy, the contribution of the MDM is dominant, and the electric quadrupole moment can be neglected ($\gamma(q^2)=0$).\\

We define the notation for the process as $ e^{+}(k_+) e^{-}(k_-) \to K^+(p_1) \pi^0(p_2) K^-(p_3) \pi^0(p_4)$, where the quantities in parentheses denote the corresponding particle four-momenta. The total amplitude can be written in the generic form
\begin{equation}
\mathcal{M}=\frac{e}{q^2} l^\mu J_\mu,
\label{Eq:amplitude}
\end{equation}
where the factor $1/q^2$ comes from the photon propagator ($q=k_+ +k_-$). The leptonic current $l^{\mu} \equiv  \bar{v}(k_+) \gamma^\mu u(k_-)$ is common to all channels, while $J_\mu$ represents the hadronic electromagnetic current, which includes the contributions of all considered channels
\begin{equation}
    J^\mu =\sum_\text{channel}  \mathcal{M}^{\mu}_\text{channel}.
\end{equation}
This current must satisfy the Bose-Einstein symmetry, by the exchange of neutral pions, and the $C$ invariance, by the exchange of charged kaons.
The contribution of each channel to the electromagnetic current $J^\mu$ is denoted by $\mathcal{M}^{\mu}_{\text{channel}}(p_1,p_2,p_3,p_4)$. Additional amplitudes corresponding to diagrams that differ only by mesons exchange will be represented by the appropriate exchange of momenta  in the corresponding function.

The $\gamma^* \to 2K2\pi$ vertex is described by considering the relevant hadronic degrees of freedom in the energy range of interest, guided by the BaBar analysis \cite{BaBar:2011btv} and modeled within the VMD approach \cite{Kroll:1967it,Bando:1984ej,Fujiwara:1984mp,Meissner:1987ge}. 
We consider the effective Lagrangian for the Vector-photon and vector-pseudoscalar-pseudoscalar interaction
\begin{eqnarray}
{\cal L}&=&
\sum_{V} \frac{e\, m_V^2}{g_V}\,V_\mu\, A^\mu + g_{VP_1P_2}\,
\, V_\mu\, (P_1 \,\partial^\mu\, P_2 - P_2\,\partial^\mu\, P_1),\nonumber\\
\label{Lvmd}
\end{eqnarray}
where $V$ and $A$ refer to the vector meson and the photon fields, respectively, while $P_i$ are pseudoscalar fields. The couplings are treated as free parameters that can be determined from experimental information. 

\begin{figure}
\begin{center}
\includegraphics[scale=0.35]{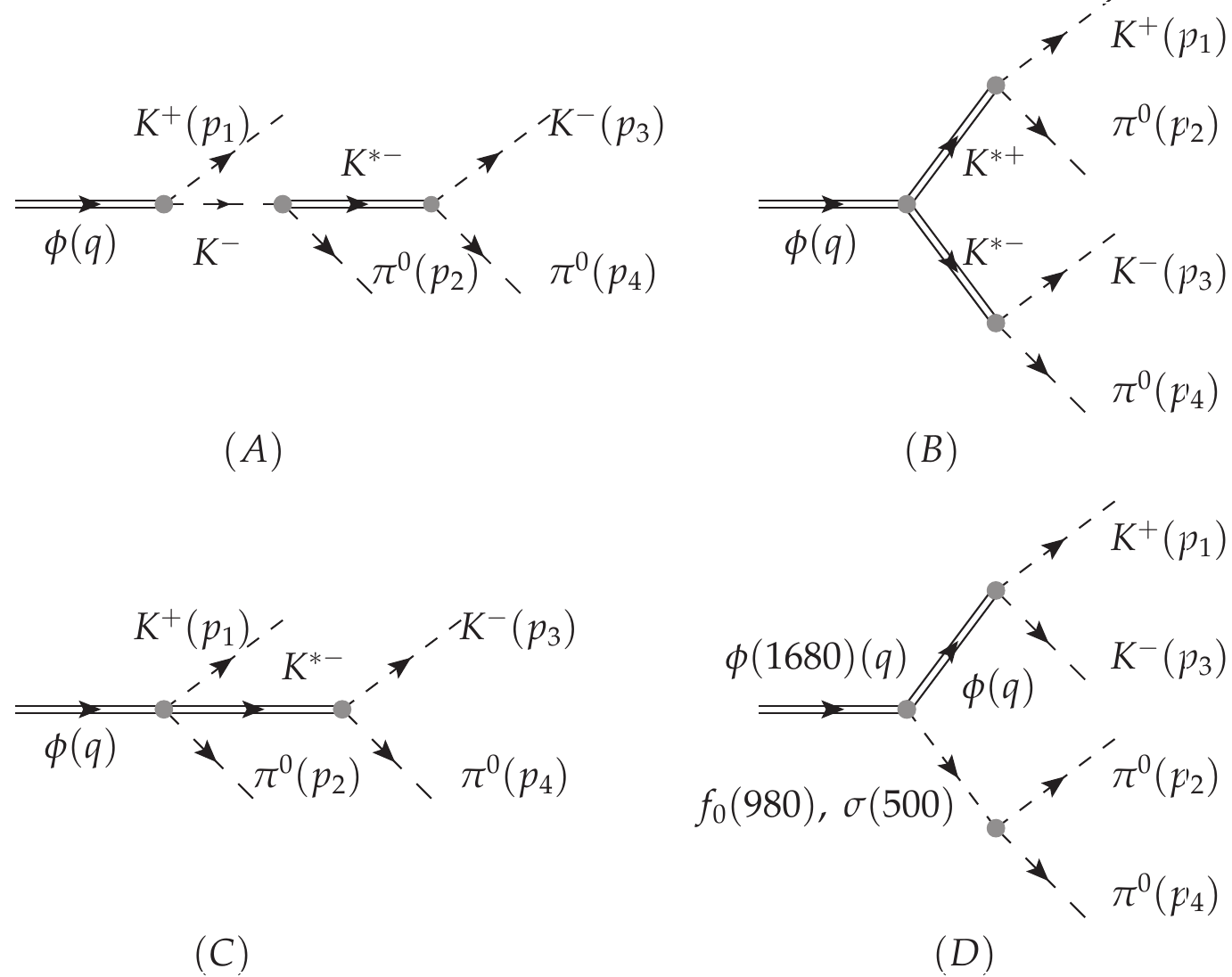}
\end{center}
\caption{Generic channels considered for the description of the $e^{+} e^{-} \to K^+ K^- 2 \pi^0$  process. The full set of diagrams is obtained by applying Bose-Einstein symmetry and charge conjugation to each channel. Channels $A$, $B$ and $C$ are linked by gauge invariance, while channel $D$ is gauge invariant by itself.}
\label{fig:channels}
\end{figure}

The channels considered in this work are shown in Fig. \ref{fig:channels}. 
In all channels, the initial vector state is coupled to the virtual photon from the leptonic current. We provide the hadronic part of the amplitude of each channel excluding the common photon propagator, consistent with Eq. \ref{Eq:amplitude}. We discuss them below.\\

{\it Channel $A$}\\
The probability amplitude of channel $A$ is given by
\begin{eqnarray}
     \mathcal{M}^{\mu}_{A}(p_1,p_2,p_3,p_4) = -e\,
 \frac{g_{\phi K K}g^2_{K^* K\pi}}{g_{\phi}}\,m^{2}_{\phi}\,D^{\mu\alpha}_{\phi}[q]\,\nonumber\\
 \times(q-2p_1)_{\alpha}\,S_{K}[q-p_1](q-p_1+p_2)_{\gamma}\,D^{\gamma\eta}_{K^{*}}[s_{34}]\,r_{43\eta},\nonumber\\
 \label{MA1}
\end{eqnarray}
where we have defined $s_{ij}\equiv p_i + p_j$, $ r_{ij} \equiv  p_i-p_j$. $S_{K}[q]\equiv i/(q^{2}-m^{2}_{K}) $
is the kaon propagator with mass $m_{K}$. The propagator of vector mesons, $D^{\alpha\mu}_V[p]$, with mass $m_V$ and decay width $\Gamma_{V}$, is assumed to have the complex mass form: 
\begin{equation}
    D^{\alpha\mu}_V[p]=-iD_V[p] \left(g^{\alpha\mu}-\frac{p^{\alpha}p^{\mu}}{m^{2}_V-i\,m_V\Gamma_V}\right),
\end{equation} 
where $D_V[p]\equiv 1/(p^{2}-m^{2}_V + i\,m_V\Gamma_V)$. 
This form is consistent with the inclusion of the absorptive corrections and the requirement of gauge invariance \cite{Baur:1995aa,Argyres:1995ym,Beuthe:1996fe,LopezCastro:1999xg}. In some cases, the label referring to the vector meson includes the corresponding charge to facilitate the identification of the particle in the corresponding Feynman diagram.
The main decay mode of the $K^*$ meson is in a pair $K\pi$, the energy dependence of the width is 
\begin{equation}
\Gamma_{K^*}(s)=\Gamma_{K^*} \left(\frac{m_{K^*}}{\sqrt{s}}\right)^5\,\left[\frac{\lambda(s,m_K^2,m_\pi^2)}
{\lambda(m_{K^*}^2,m_K^2,m_\pi^2)}\right]^{3/2},
\end{equation}
where $\lambda(x,y,z)$ is the K\"all\'en function and $\Gamma_{K^*}$ is the decay width at $\sqrt{s}=m_{K^*}$. The amplitude, Eq. (\ref{MA1}), can be written in the simplified form
\begin{eqnarray}
    \mathcal{M}_{A}^{\mu}(p_1,p_2,p_3,p_4) &=& i\,e\,C_A\, D_{K^{*}}[s_{34}]\dfrac{x_1^\mu}{x_1\cdot q} \nonumber\\
    &\times&\left( r_{43} \cdot  z_{12} -\bar\Delta[s_{34}]\,z_{12} \cdot s_{34}\right),
    \label{MAr}
\end{eqnarray}
where $x_1 \equiv q-2\,p_1  $, $ z_{12} \equiv q-p_1+p_2$ and
\begin{eqnarray}
C_A&\equiv& (g_{\phi\,K\, K}\,g^2_{K^* K \pi}/g_{\phi})\,m^{2}_{\phi} D_{\phi}[q], \nonumber\\
 \Delta &\equiv& m_{\pi}^2 - m_{K}^2,\nonumber\\
dK^*[s_{34}] &\equiv& m^2_{K^{*}} - im_{K^{*}}\Gamma_{K^{*}}(s_{34}),\nonumber\\
\bar\Delta[s_{34}] &\equiv& \Delta/dK^*[s_{34}]. \nonumber \end{eqnarray}

\vspace*{.5cm}
{\it Channel B}\\
In Fig. \ref{fig:channels} ($B$), we show the so-called $K^*$-channel, where the $\phi$ meson couples to a $K^*$ pair, through a triple vector vertex. This defines the $K^*$ electromagnetic vertex, times a global constant $g_{\phi K^* K^* }$ that accounts for the intermediate strong interaction process. The corresponding amplitude is given by
\begin{align}
\mathcal{M}^{\mu}_{B}(p_1,p_2,p_3,p_4) =& 
e\,\frac{g_{\phi K^{*}K^{*}}\, g^2_{K^{*}K\pi}}{g_{\phi}}\,m^{2}_{\phi}\,D^{\mu\alpha}_{\phi}[q]r_{21\gamma}\nonumber\\
&D^{\gamma\lambda}_{K^{*+}}[s_{12}]\,\Gamma^{1}_{\alpha\lambda\delta}\,D^{\delta\eta}_{K^{*-}}[s_{34}]\,r_{43\eta}
      \label{MB1}
\end{align}
where $\Gamma^{1}_{\alpha\lambda\delta} \equiv (1+i\Gamma_V/M_V)\,\Gamma_{\alpha\lambda\delta}$ represents the absorptive corrected vertex at one-loop level, consistent with gauge invariance \cite{Baur:1995aa,Argyres:1995ym,Beuthe:1996fe,LopezCastro:1999xg}. The tree-level vertex, Eq.~(\ref{vertex}), for this momentum configuration takes the form
\begin{flalign}
      \Gamma_{\alpha\lambda\delta} &= 
      g_{\lambda\delta}\,\,Q_{1}{}_{\alpha} + \beta\,(q_{\delta}\,\,g_{\alpha\lambda}-q_{\lambda}\,\,g_{\delta\alpha}) \nonumber\\
      &\,\,\,\,\,\,+ s_{12\lambda}\,\,g_{\delta\alpha} - s_{34\delta}\,\,g_{\alpha\lambda}\nonumber,
\label{G0vertex}
\end{flalign}
where $q = s_{12} + s_{34}$ and $Q_{1} \equiv s_{34}-s_{12}$, and we have omitted the dependence on $q^2$ of $\beta$ since the energy region to scan is small and therefore can be considered constant. The amplitude can be written in a simplified form
\begin{equation}
 \begin{aligned}
	&\mathcal{M}_B^\mu(p_1,p_2,p_3,p_4)=\\
    &i\,e\,C_B \dfrac{Q_1^\mu}{q\cdot Q_1}\left[D_{K^{*}}[s_{34}]\left(r_{21}\cdot s_{34}\,\bar\Delta[s_{34}] - r_{21}\cdot r_{43}\right)\right.\\
	&- \left.D_{K^{*}}[s_{12}]\left(r_{43}\cdot s_{12}\,\bar\Delta[s_{12}] - r_{21}\cdot r_{43}\right)\right]\\
    &+ \mathcal{M}_B^\mu(p_1,p_2,p_3,p_4)_{GI}
    \label{MBr}
 \end{aligned}
\end{equation}
where $C_B = \frac{g_{\phi K^{*}K^{*}}g^2_{K^* K \pi}}{g_\phi}m^{2}_{\phi} D_{\phi}[q]$, and we define the gauge invariant tensor
\begin{equation}
    L^{\mu}(a,b) \equiv \frac{a^{\mu}}{a\cdot q} - \frac{b^{\mu}}{b\cdot q}.
    \label{Lmunu}
\end{equation}
$\mathcal{M}_B^\mu(p_1,p_2,p_3,p_4)_{GI}$ is the self gauge invariant part of channel $B$, and explicitly given in Appendix B.\\

{\it Channel C}\\
In Fig. \ref{fig:channels} ($C$), we show the process driven by a contact term ($\phi K^* K \pi$), whose amplitude can be written in the general form 
\begin{eqnarray}
     \mathcal{M}^{\mu}_{C}(p_1,p_2,p_3,p_4)=i\,e\, \frac{g_{\phi K^* K \pi}\,g_{K^* K \pi}}{g_{\phi}}m^{2}_{\phi}\nonumber\\
     \times D^{\mu\alpha}_{\phi}[q]\,\Gamma^{1}_{\alpha\gamma}\,D_{K^{*}}^{\gamma\eta}[s_{34}]\,r_{43\eta}.
     \label{MCr}
\end{eqnarray}

Channels $A$, $B$ and $C$, together with their corresponding neutral and charged kaons exchanges, are linked by the gauge invariance condition. The effective coupling $g_{\phi K^* K \pi}$ and the general vertex $\Gamma^{1}_{\alpha\gamma}$ are fixed by such a condition. This is an important observation, since our description of channels $A$ and $B$ is not fully gauge invariant by itself. Here, we describe how the three channels are properly combined in order to obtain an analytically gauge invariant amplitude. To this end, we profit from the Ward-Takahashi identity, fulfilled by the $VV\gamma$ vertex, instead of introducing the counter-term as a general requirement. This procedure allows us to trace the origin of the different contributions. The details are presented in Appendix B.

In order to obtain the gauge invariant amplitude, we use the following combination of the amplitudes
\begin{eqnarray}
\mathcal{M}^{\mu}_{ABC_{24}} &=& \mathcal{M}^{\mu}_{A}(p_1,p_2,p_3,p_4)
+\mathcal{M}^{\mu}_{A}(p_3,p_4,p_1,p_2)\nonumber\\
&&+\mathcal{M}^{\mu}_{B}(p_1,p_2,p_3,p_4)\\
&&+\mathcal{M}^{\mu}_{C}(p_1,p_2,p_3,p_4)+\mathcal{M}^{\mu}_{C}(p_3,p_4,p_1,p_2),\nonumber
\end{eqnarray}
where we have adopted a notation that refers to the three channels and the momenta of the neutral pions.
The resulting gauge invariant amplitude is
\begin{equation}
    \begin{aligned}
    &\mathcal{M}_{ABC_{24}}^\mu=\\
    &ieC\left\{L^\mu(x_1,x_3)\left[D_{K^{*}}[s_{34}]\left(r_{43}\cdot z_{12}-z_{12}\cdot s_{34}\,\bar \Delta[s_{34}]\right)\right. \right.\\
	&\left.+D_{K^{*}}[s_{12}]\left(r_{21}\cdot z_{34} - z_{34}\cdot s_{12}\,\bar\Delta[s_{12}]\right)\right]\\
	&+L^\mu(Q_1,x_3)D_{K^{*}}[s_{34}]\left(-r_{21}\cdot r_{43} + r_{21}\cdot s_{34}\,\bar\Delta[s_{34}]\right)\\
	&\left.-L^\mu(Q_1,x_1)D_{K^{*}}[s_{12}]\left(-r_{21}\cdot r_{43}+r_{43}\cdot s_{12}\,\bar\Delta[s_{12}]\right)\right\}\\
    &+ \mathcal{M}_B^\mu(p_1,p_2,p_3,p_4)_{GI}, 
    \label{abc24}
    \end{aligned}
\end{equation}
where $C \equiv C_A=C_B= (g_{\phi\,K\, K}\,g^2_{K^* K \pi^0}/g_{\phi})\,m^{2}_{\phi}\,D_{\phi}[q]$ are required, $x_1 \equiv q-2\,p_1$ and $x_3 \equiv q-2\,p_3$. A similar expression is obtained by adding the remaining amplitudes
\begin{eqnarray}
 \mathcal{M}^{\mu}_{ABC_{42}} &=& \mathcal{M}^{\mu}_{A}(p_1,p_4,p_3,p_2)+\mathcal{M}^{\mu}_{A}(p_3,p_2,p_1,p_4)\nonumber\\
 &&+\mathcal{M}^{\mu}_{B}(p_1,p_4,p_3,p_2)\label{abc42}\\
 &&+\mathcal{M}^{\mu}_{C}(p_1,p_4,p_3,p_2)+\mathcal{M}^{\mu}_{C}(p_3,p_2,p_1,p_4),\nonumber
\end{eqnarray}
which in practice corresponds to the exchange $p_2 \leftrightarrow p_4$.\\

{\it Channel D}\\
To model the remaining part of the electromagnetic current, following the analysis performed by BaBar, we consider the channel in which the photon couples the $\phi(1680)$ vector meson, which subsequently decays into the $\phi$ meson and the $\sigma$ or $f_0(980)$ scalar meson, as shown in Fig. \ref{fig:channels}($D$). The effective Lagrangian describing the $VVS$ and $PPS$ interactions (where $V$, $P$, and $S$ denote vector, pseudoscalar, and scalar fields, respectively) is given by the following
\begin{eqnarray}\label{LD}
    \mathcal{L}_S &=& ig_{V_1 V_2 S}\, (\partial_\beta V^{\mu}_{1}\,\partial^\beta V_{2\mu}\, -\partial_\beta V^{\mu}_{1}\,\partial_\mu V^{\beta}_{2}\,)S\nonumber\\
    &&+ ig_{SPP}\,S\,P_1\,P_2,
\end{eqnarray}
where $g_{V_1V_2 S}$ and $g_{S P_1 P_2}$ are effective coupling constants. The simplified amplitude, considering the contribution of $f_0(980)$, is given by
\begin{eqnarray}
	\mathcal{M}^\mu_{D_{f_0}}\left(p_1,p_2,p_3,p_4\right) &=& -ieC_{f_0} D_{\phi}[s_{13}]D_{f_0}[s_{24}]\nonumber\\
	&&\times  \dfrac{F^{\mu\beta}(s_{13},q)}{q\cdot s_{13}}r_{31\beta},
\end{eqnarray}
where the propagator of the $f_0(980)$ meson is $D_{f_0}[s]\equiv i/(s-m_{f_0}^2+i m_{f_0}\Gamma_{f_0})$ and $	C_{f_0} \equiv (g_{f_0\pi^0\pi^0}\, g_{\phi K^+K^-}\,  g_{\phi'\phi f_0}/g_{\phi'})m^2_{\phi'}D_{\phi'}[q]$, with $\phi'=\phi(1680)$ and the tensor function is given by
\begin{equation}
\begin{aligned}
 F_{\mu\alpha}(a,b) \equiv a\cdot b\,\,g_{\mu\alpha}-a_{\mu}b_{\alpha}.
 \label{LamOme1}
\end{aligned}
\end{equation}

\noindent Notice that this channel does not involve additional exchange of neutral or charged pions, since both pions are emitted from the same vertex. Analogous expression is obtained for the $\sigma$ meson. 
The combined amplitude for the $f_0(980)$ and $\sigma$ mesons is
\begin{equation}
\begin{aligned}
&\mathcal{M}^{\mu}_{D}\left(p_1,p_2,p_3,p_4\right)=\nonumber\\
&\mathcal{M}^{\mu}_{D_{f_0}}(p_1,p_2,p_3,p_4)
+e^{i\theta_1}
\mathcal{M}^{\mu}_{D_{\sigma}}(p_1,p_2,p_3,p_4),
\end{aligned}
\end{equation}
where the two contributions are combined with a relative phase $\theta_1$. The decay width of the particles involved in this channel is considered as constant. This description is sufficient to describe the currently available low precision data.

Finally, the total hadronic electromagnetic current can be written as
\begin{equation}
J^{\mu}=
\mathcal{M}^{\mu}_{ABC}
+e^{i\theta_2}
\mathcal{M}^{\mu}_{D},
\end{equation}
where $\mathcal{M}^{\mu}_{ABC}$ includes the contributions ${ABC_{24}}$ and ${ABC_{42}}$. A relative phase $\theta_2$ between the two amplitudes is introduced as a free parameter.

\section{The $K^*$ MDM from the total cross section}	
We compute the total cross section following the PDG convention \cite{pdg}, neglecting the electron mass
\begin{equation}
    d\sigma = \frac{(2\pi)^{4}\,\overline{|\mathcal{M}|^{2}}}{4\,\sqrt{(k_+\cdot k_-)^{2}}}\,\delta^{4}\Big(q-\sum^{n}_{i=1}p_i\Big)\prod^{n}_{i=1}\frac{d^{3}p_i}{(2\pi)^{3}\,2E_i}.\label{sigmaee4pi2}
\end{equation}
The spin-averaged squared amplitude  $|\overline{\mathcal{M}|^{2}}$ is built up by the leptonic and hadronic contributions discussed above
\begin{equation}
    \overline{|\mathcal{M}|^{2}} = \frac{e^2}{q^{4}}\,l_{\mu\nu}\,h^{\mu\nu}, 
\label{BarM}
\end{equation}
where $h^{\mu\nu} \equiv J^{\mu}\,J^{\nu\dagger}$ is the hadronic tensor constructed from the hadronic currents discussed above, and $l_{\mu\nu}$ is the leptonic tensor, given by
\begin{equation}
    l_{\mu\nu} = k_{+\mu}\,k_{-\nu}+k_{-\mu}\,k_{+\nu}-\frac{q^{2}}{2}\,g_{\mu\nu}.
\label{lmunu2}
\end{equation}

The kinematical variables and integration limits are chosen following Ref.\cite{Kumar:1969jjy}. The integration is performed numerically using a Fortran code and the Vegas subroutine \cite{PETER1978192}.

For the numerical analysis, we use $m_{f_0}=0.972 \pm 0.002$ GeV and $\Gamma_{f_0}=0.056 \pm 0.011$ GeV, obtained by BaBar \cite{BaBar:2011btv} from the same data set. The coupling $g_{f_0\pi^0\pi^0} = 1.0\pm0.1$ GeV is determined from the decay width of $f_0 \to \pi\pi$, where the two pion mode accounts for the total decay width.
Similarly, the coupling $g_{\sigma\pi^0\pi^0} $ is determined from the corresponding decay width of the $\sigma$ meson which, together with its mass, is fixed by fitting to the cross section. See appendix A for further details on the couplings involved.

We perform a fit to the cross section experimental data using the Minuit minimization subroutine. In Table \ref{tab:freeparam} we list the free parameters and their numerical values obtained from the fit.

\begin{table}[h]
    \centering
    \begin{tabular}{|c|c|}
        \hline
        Parameter & Description \\
        \hline \hline
        $\theta_1=2.9 \pm 0.29 (\pi\, \text{units})$ & $\sigma$ to $f_0(980)$  phase \\
        \hline
        $\beta=1.5 +1.2$ & Magnetic dipole moment  \\
        \hline
        $\theta_2=1.2 \, (\pi\, \text{units})$ & $D$ to $A,B,C $ phase \\
        \hline
        $g_{\phi^\prime\phi f_0}/g_{\phi^\prime}=1.15^{+0.07}_{-0.15}$ & Couplings ratio\\
        \hline
        $g_{\phi^\prime\phi \sigma}/g_{\phi^\prime}=0.32 \pm 0.06$ & Couplings ratio\\
        \hline
        $m_\sigma=0.71\pm0.02$ (GeV) & $\sigma$-meson mass\\
        \hline
        $\Gamma_{\sigma}=0.13^{+0.13}_{-0.08}$ (GeV) & $\sigma$-meson width\\
        \hline
    \end{tabular}
    \caption{Free parameters. The values were obtained from a fit to data using Minuit. See text for details.}
    \label{tab:freeparam}
\end{table}

\begin{figure}[t]
\begin{center}
\includegraphics[scale=0.3]{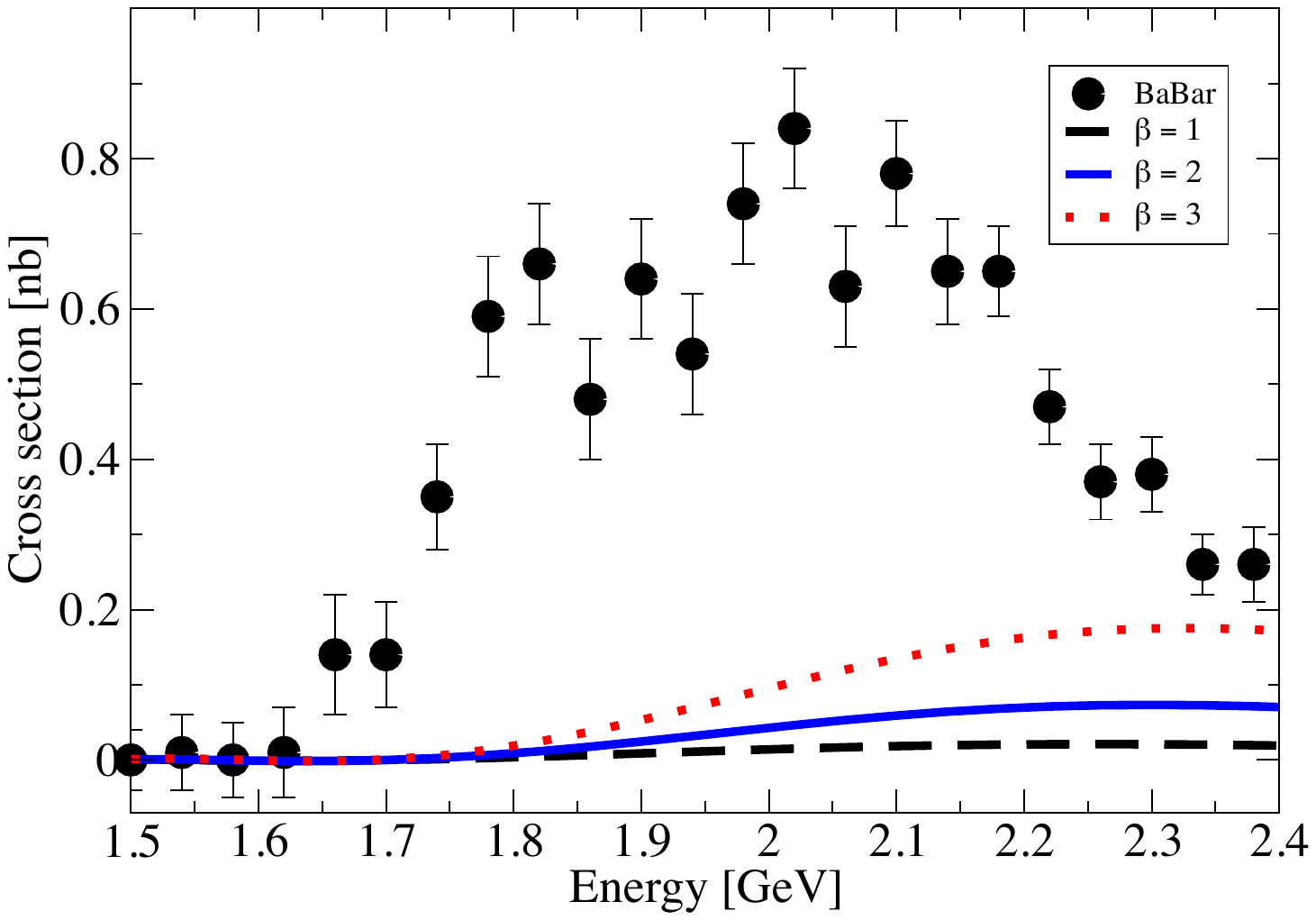}
\end{center}
\caption{ $\beta$ dependence on $e^{+}\,e^{-}\to K^+ K^- \, 2\pi^{0}$ cross section. BaBar data (symbols) and the contribution from $A$, $B$ and $C$ channels (together) for $\beta =1,\ 2, \ 3$ (dashed, solid, and dotted lines, respectively).}
\label{betasense}
\end{figure}

In order to illustrate the dependence on the parameter $\beta$, in Fig. \ref{betasense}, we show the cross section contribution from $A+B+C$ channels. We consider $\beta =$1, 2, and 3. These results are compared with the experimental data from BaBar. We observe that the dependence on $\beta$ is quadratic, causing the contribution to the cross section to increase rapidly, as does $\beta$. This feature will be important for establishing an upper bound. The plots also show that the main sensitivity to $\beta$ occurs in the region between 1.8 GeV and 2.4 GeV. For energies above 2.4 GeV additional structures appear that are not captured by the present description, and therefore we limit our fit to energies below 2.4 GeV.

\begin{figure}[t]
\begin{center}
\includegraphics[scale=0.3]{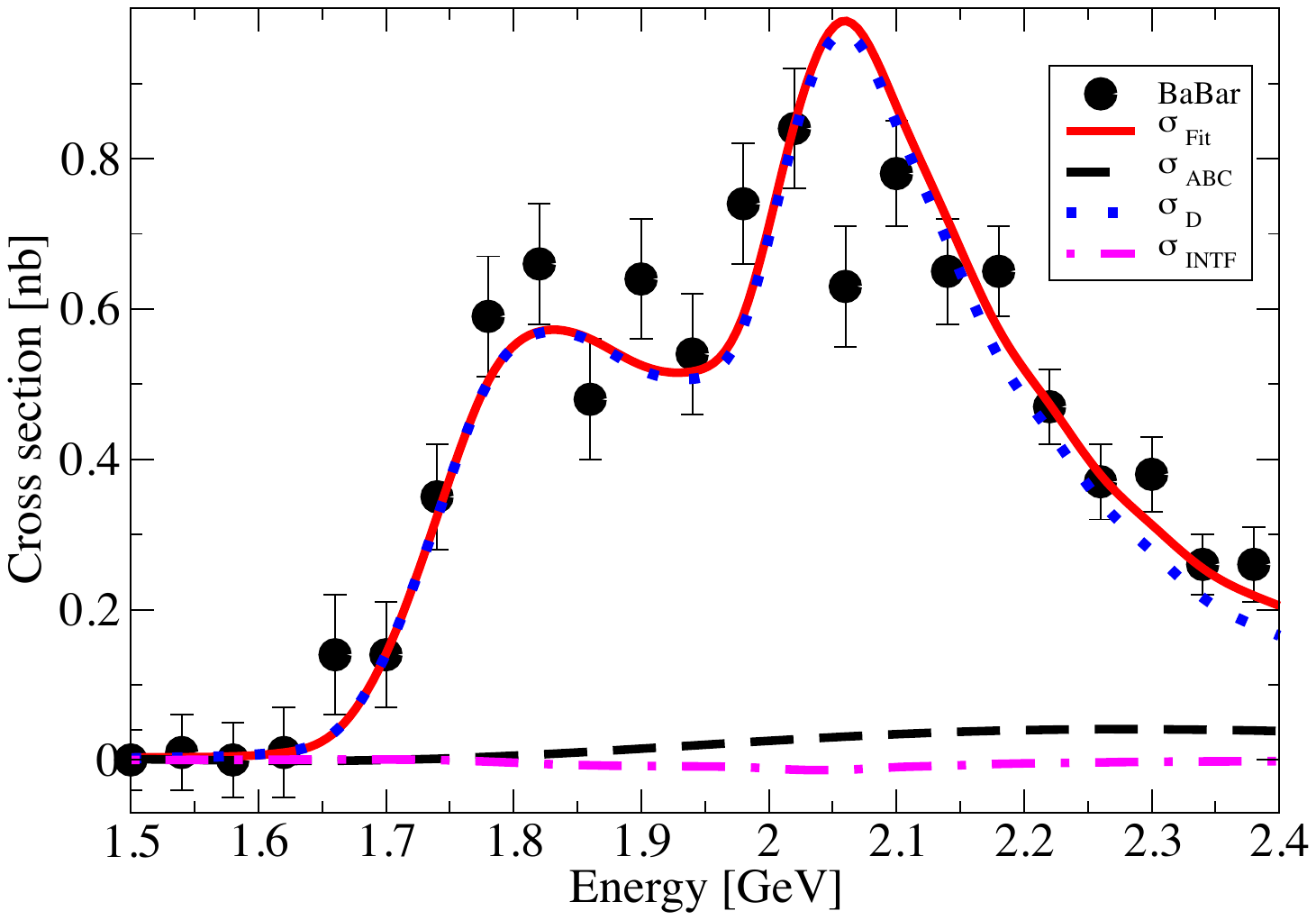}
\end{center}
\caption{$e^{+}\,e^{-}\to 2\pi^{0}\pi^{+}\pi^{-}$ cross section. BaBar data (symbols) and the fit to the total cross section (solid line). We also show the contribution of each channel to the total cross section: $A$, $B$ and $C$ channels grouped together (dashed line), $D$ (dotted line), and the interference between them (dot-dashed line).}
\label{allchannels}
\end{figure}

In Fig. \ref{allchannels}, we show the total cross section data from BaBar \cite{BaBar:2011btv} (symbols) and the corresponding fit from this work (solid line). In the same Figure, we show the contribution of each channel: $A$, $B$ and $C$ channels grouped together (dashed line), $D$ (dotted line), as well as the interference between channels $ABC$ and $D$. Error bars are not displayed for the sake of clarity. We observe that the interference between the $ABC$ and $D$ channels is very small. For this reason, the relative phase ($\theta_2$) between these channels cannot be determined with sufficient precision. Therefore, in Table \ref{tab:freeparam}, we quote only a central value favored by the fit, which does not significantly modify the final result.\\

\begin{figure}[t]
\begin{center}
\includegraphics[scale=0.3]{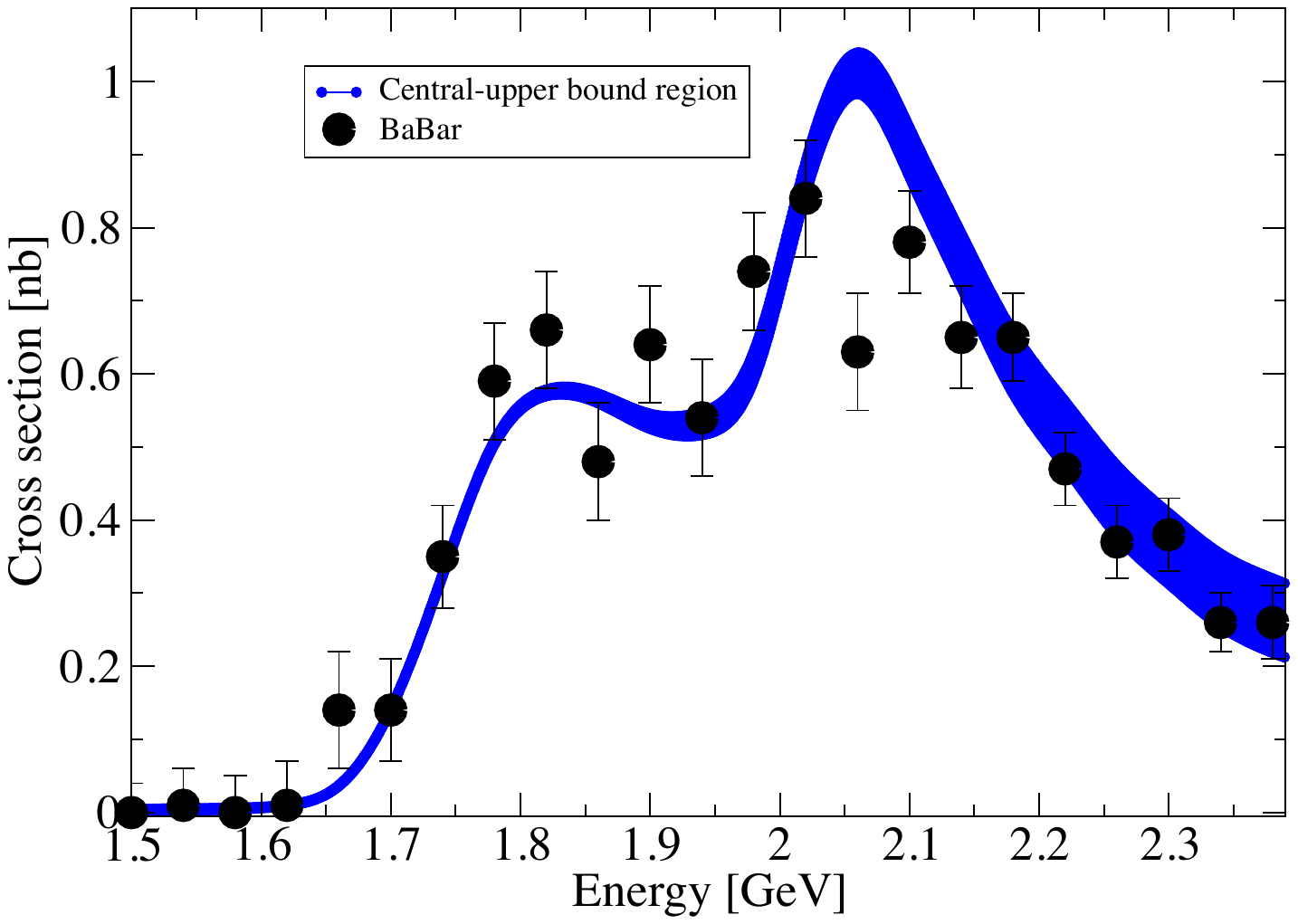}
\end{center}
\caption{BaBar data for the $e^{+}\,e^{-}\to K^+ K^- \, 2\pi^{0}$ total cross section (symbols with error bars). The shadowed region is delimited by the central value and the upper bound of the $K^*$ MDM.}
\label{total}
\end{figure}

The fit favors $\beta= 1.5 + 1.2$ with $\chi^2/dof= 1.6$. The upper value was determined by identifying the increase in $\chi^2/dof$ by one unit. No lower bound is quoted, as the experimental data are unable to constrain this parameter. In order to properly obtain the static limit value of the MDM, we consider the static limit of the electric charge form factor
\begin{equation}
    |F_{K^* }\left(0\right)| =
\lim_{q^2 \to 0} 
\left\lvert \frac{g_{\phi KK}m_{\phi}^2 }{g_{\phi}}D_{\phi}[q^2]
\right\rvert=1.\nonumber
\label{fformarec}
\end{equation}
Using the value of the parameters, we obtain $|F_{K^* }\left(0\right)|=0.33 \pm 0.005$, where the uncertainty arises mainly from the  uncertainty in $g_\phi$. In order to properly define the MDM extracted, we normalize $\beta$ to this value. Thus, the central value of the MDM of the $K^*$ meson, extracted from the BaBar data, is  
\begin{equation}
\mu_{K^*} = 4.5 \ \ \text{in units of } \ (e/2 m_{K^*}),
\label{resfinbeta}
\end{equation}
with an upper bound of $\bar{\mu}_{K^*} = 6.3$ in units of $(e/2 m_{K^*})$.\\ 
In Fig. \ref{total}, we show the experimental cross section as measured by BaBar (symbols with error bars). The shadowed region is delimited by the central value and the upper bound of the $K^*$ MDM. 

\section{Conclusions}
We have performed an analysis of the $e^+ e^- \to K^+ K^- 2 \pi^0$ cross section measured by the BaBar Collaboration \cite{BaBar:2017zmc}, to determine its sensitivity to the MDM of the $K^*$ meson. We modeled the process in the VMD approach, obtaining explicit gauge invariant expressions for the corresponding amplitudes, in particular for the $A$, $B$, and $C$ channels. This result generalizes the analogous result obtained for the $\rho$ meson \cite{Rojas:2024tmn}, incorporating the isospin symmetry breaking contribution.

We have shown that the cross section is sensitive to the MDM of the $K^*$  and that it may provide the possibility to determine a first value based on experimental data. The outcome of the numerical analysis favors the central value $\mu_{K^*} = 4.5$ with an upper bound of $\mu_{K^*} < 6.3 \ \ \text{in units of} \ (e/2 m_{K^*}) $. This result is limited by the currently available low precision data. We consider it important to stress the need to measure this process with higher precision, which could provide a determination of this fundamental property of the $K^*$ meson and contribute to a deeper understanding of the QCD structure of hadrons.

\appendix
\section{Couplings}
For consistency, we provide the general expressions used to evaluate the couplings involved in this work. These couplings have been discussed in more detail in Ref.~\cite{Avalos:2022txl} within the same approach. The values of the parameters involved are taken from the PDG \cite{pdg}, unless otherwise stated.\\

$VP_1P_2$ {\it coupling}\\

The coupling constant associated with the vertex $VP_1 P_2$ in the decay $V(q)\to P_1(p_1)P_2(p_2)$ is given by
\begin{equation}
    g_{V P_1 P_2}=\left(\dfrac{48\,\pi\, m_V^5\,\Gamma(V\to P_1 P_2)}{\lambda^{3/2}(m_V^2,m_{P_1}^2,m_{P_2}^2)}\right)^{1/2}.
\end{equation}
For the decay $\phi \to K^+ K^-$, we obtain $g_{\phi K K}= 4.47 \pm 0.03$.\\
\noindent For the decay $K^{*+}\to K^+\pi^0$, we obtain
$g_{K^{*} K \pi}=3.05\pm 0.05,$
where we make use of the isospin relation,
\begin{equation}
    \Gamma(K^{*+}\to K^+\pi^0)=\dfrac{1}{3}\Gamma(K^{*+}\to K\pi).
\end{equation}

$V\gamma$ {\it coupling}\\

This coupling is determined from the leptonic decay $V\to l^+l^-$, within the VMD approach
 \begin{equation}
     g_V = \sqrt{\dfrac{4\pi\alpha^2\left(2m_l^2+m_V^2
     \right)\sqrt{m_V^2-4m_l^2}}{3m_V^2\Gamma(V\to l^+l^-)}}.
 \end{equation}
In the case of interest for this work, the process is $\phi\to e^+e^-$. We obtain
\begin{equation*}
	g_{\phi }= 13.40 \pm 0.10.
\end{equation*}

$SPP$ {\it coupling}\\

The coupling constant $g_{SPP}$ is calculated using 
\begin{equation}
    g_{SPP} = \sqrt{\dfrac{16\,\pi\,M^2\,\Gamma\left(S\to PP\right)}{\sqrt{M^2-4m^2}}},
\end{equation}
where $M$ ($m$) denotes the mass of the scalar (pseudoscalar) particle, and $\Gamma$ is the  total decay width.

In this work, we have two processes of interest: $f_0 \to \pi^0 \pi^0$ and $\sigma \to \pi^0 \pi^0$. For the $f_0$ meson, we use $M_{f_0} = 0.972 \pm 0.002$ and $\Gamma_{f_0} = 0.056 \pm 0.011$ GeV, as obtained from the BaBar Collaboration from the same dataset under consideration \cite{BaBar:2011btv}.
The total
and partial decay widths are related by
\begin{equation}
	\Gamma\left(f_0\to\pi^0\pi^0\right) = \dfrac{1}{3}\Gamma\left(f_0\to\pi\pi\right).
\end{equation}
Hence, the value of the coupling constant $g_{f_0\pi^0\pi^0}$ is
\begin{equation}
	g_{f_0 \pi^0\pi^0} = 1.0\pm 0.1 \ \ \text{GeV}.
\end{equation}
For the decay $\sigma \to \pi^0\pi^0$ we use the same relations, with the mass and width of the $\sigma$ meson determined from the fit to the cross section and
\begin{equation}
	g_{\sigma \pi^0\pi^0} =  1.3\pm 0.5 \text{GeV}.
\end{equation}

\section{Channels $A$, $B$ and $C$ gauge invariant amplitude including isospin symmetry breaking}
We derive the fully gauge invariant amplitude for channels $A$, $B$, and $C$.
The condition of gauge invariance requires
\begin{equation}
q_{\mu}\,\Big(\mathcal{M}^{\mu}_{A} + \mathcal{M}^{\mu}_{B}\Big) = -q_{\mu}\,\mathcal{M}^{\mu}_{C}. 
\end{equation}
The amplitude $\mathcal{M}^{\mu}_{A}(p_1,p_2,p_3,p_4)$, Eq.~(\ref{MAr}), upon contraction with $q_{\mu}$, becomes
\begin{eqnarray}
    q_{\mu}\,\mathcal{M}^{\mu}_{A}(p_1,p_2,p_3,p_4) &=& i\,e\, C_A D_{K^{*}}[s_{34}]\nonumber\\
    &&\times \,\Big[z_{12}\cdot r_{43} - z_{12}\cdot s_{34}\,\bar\Delta[s_{34}]\Big],\nonumber\\
    \label{qMA1}
\end{eqnarray}
where $x_1 \equiv q-2\,p_1$, $z_{12} \equiv q-p_1+p_2$ and $q=s_{21}+s_{43}$.\\
In a similar way, the amplitude $\mathcal{M}^{\mu}_{B}(p_1,p_2,p_3,p_4)$, Eq.~(\ref{MBr}), upon contraction with $q_{\mu}$, becomes
	\begin{eqnarray}
    q_\mu\mathcal{M}_B^\mu&&(p_1,p_2,p_3,p_4)=\nonumber\\
    &&i\,e\,C_B\Big[D_{K^*}[s_{34}]\left(r_{21}\cdot s_{34}\,\bar\Delta[s_{34}] - r_{21}\cdot r_{43}\right)\nonumber\\
	&& -D_{K^*}[s_{12}]\left(r_{43}\cdot s_{12}\,\bar\Delta[s_{12}] - r_{21}\cdot r_{43}\right)\Big].
    \label{qMB1}
	\end{eqnarray}

where the gauge invariant part of channel $B$ is

\begin{equation}
 \begin{aligned}
 &\mathcal{M}_B^\mu(p_1,p_2,p_3,p_4)_{GI}=\\
   & i\,e\,C_B\,(1+i\gamma)D_{K^{*}}[s_{12}]D_{K^{*}}[s_{34}]\Big[\\
    &\Delta\left\{\left(\dfrac{s_{12}^2}{dK^*[s_{12}]} -1\right)\right.\\
	&\times\left[q\cdot r_{43}L^\mu(Q_1,r_{43}) - \bar\Delta[s_{34}]\,q\cdot s_{34}L^\mu(Q_1,s_{34})\right]\\
	&-\left(\dfrac{s_{34}^2}{dK^*[s_{34}]} - 1\right)\\
    &\times\left.\left[q\cdot r_{21}L^\mu(Q_1,r_{21}) - \bar\Delta[s_{12}]\,q\cdot s_{12}L^\mu(Q_1,s_{12})\right]\right\}\\
    &+\beta(q^2) \Big\{q\cdot r_{43}\,q\cdot r_{21}\,L^\mu(r_{21},r_{43})\\
    &-\bar\Delta[s_{34}]\left[q\cdot r_{21}\,q\cdot s_{34} \,L^\mu(r_{21},s_{34})\right]\\
    &+\bar\Delta[s_{12}]\left[ q\cdot s_{12}\,q\cdot r_{43}\,L^\mu(r_{43},s_{12})\right]\\
	&+ \bar\Delta[s_{12}]\bar\Delta[s_{34}]\,q\cdot s_{34}\,q\cdot s_{12}\,L^\mu(s_{12},s_{34})\Big\}\Big].
 \end{aligned}  
 \end{equation}
For channel $C$, Eq.~(\ref{MCr}), upon contraction with $q_{\mu}$ becomes
 \begin{eqnarray}
	q_\mu \mathcal{M}_C^\mu(p_1,p_2,p_3,p_4) &=&-ieC_C q^\alpha \Gamma_{\alpha\gamma}^1 D_{K^*}[s_{34}]\nonumber\\
    &&\times\Big(r_{43}^\gamma - s_{34}^\gamma \bar\Delta[s_{34}]\Big).
\end{eqnarray}

Therefore, by choosing the following combination of amplitudes, of $A$ and $B$ channels, we obtain

\begin{eqnarray}
	&&q_\mu\Big[\mathcal{M}^\mu_A(p_1,p_2,p_3,p_4) + \mathcal{M}^\mu_A(p_3,p_4,p_1,p_2)\nonumber\\
    &&+ \mathcal{M}_B^\mu(p_1,p_2,p_3,p_4)\Big]=\nonumber\\
	&&ieC\Big[D_{K^*}[s_{34}]\left(z_{12\gamma} - r_{21\gamma}\right)\left(r_{43}^{\gamma} - s_{34}^{\gamma} \  \bar\Delta[s_{34}]\right)\nonumber\\
	&& \qquad \ \ D_{K^*}[s_{12}]\left(z_{34\gamma} - r_{43\gamma}\right)\left(r_{21}^{\gamma} - s_{12}^{\gamma} \  \bar\Delta[s_{12}]\right)\Big].\nonumber\\
    \label{qMA14B1}
\end{eqnarray}
where $C_A = C_B \equiv C$ and
$g_{\phi K^* K^*} = g_{\phi K K}$.

Choosing the following combination of amplitudes for $C$ channel, we obtain

\begin{eqnarray}\label{qhc1234_plus_qhc3412}
	&&-q_\mu\left[\mathcal{M}^\mu_C(p_1,p_2,p_3,p_4) + \mathcal{M}^\mu_C(p_3,p_4,p_1,p_2)\right]\nonumber\\
	&& = ieC_C q^\alpha \Big[\Gamma^1_{\alpha\gamma}D_{K^*}[s_{34}]\left(r_{43}^\gamma - s_{34}^\gamma \bar\Delta[s_{34}]\right)\nonumber\\
    &&\ \ \ -\Gamma^4_{\alpha\gamma}D_{K^*}[s_{12}]\left(r_{21}^\gamma - s_{12}^\gamma \bar\Delta[s_{12}]\right)\Big]
    \label{qMC14}
\end{eqnarray}
Comparing Eqs.~(\ref{qMA14B1}) and (\ref{qMC14}), we identify $C_C = C$ and
\begin{equation}
    g_{\phi\, K^*\,K\pi} = g_{\phi\,K\,K}\cdot g_{K^*\,K\,\pi}
\end{equation}
and the corresponding structures for the vertices satisfy
\begin{equation}
 \begin{split}
  q^{\alpha}\,\Gamma^{1}_{\alpha\gamma} &= z_{12\gamma} - r_{21\gamma},\\
  q^{\alpha}\,\Gamma^{4}_{\alpha\gamma} &= z_{34\gamma} - r_{43\gamma}.
 \end{split}
\end{equation}
Factorizing $q^{\alpha}$, the vertices $\Gamma^{i}_{\alpha\gamma}$  are constructed as
\begin{equation}
 \begin{aligned}
 	\Gamma_{\alpha\gamma}^1 &= \dfrac{x_{3\alpha}}{x_3\cdot q}\left(z_{12\gamma} - r_{21\gamma}\right),\\
    \text{and}&\\
	\Gamma_{\alpha\gamma}^4 &= \dfrac{x_{1\alpha}}{x_1\cdot q}\left(z_{34\gamma} -  r_{43\gamma}\right).
 \end{aligned}
\end{equation}
Thus, the final expression for the selected amplitudes of channel $C$ is
\begin{eqnarray}
    &\mathcal{M}^{\mu}_{C}(p_1,p_2,p_3,p_4)
    =-ie  C \dfrac{x_3^\beta}{x_3\cdot q}\nonumber\\
    &\times D_{K^{*}}[s_{34}]\Big[z_{12}\cdot r_{43}-r_{21}\cdot r_{43}\nonumber\\
    &- \bar\Delta[s_{34}]\left(z_{12}\cdot s_{34} - r_{21}\cdot s_{34}\right)\Big ]\nonumber\\
\end{eqnarray}

 In order to obtain the gauge invariant amplitude, we select a particular set of amplitudes to combine , namely
 \begin{eqnarray}
\mathcal{M}^{\mu}_{ABC_{24}} &=& \mathcal{M}^{\mu}_{A}(p_1,p_2,p_3,p_4)
+\mathcal{M}^{\mu}_{A}(p_3,p_4,p_1,p_2)\nonumber\\
&&+\mathcal{M}^{\mu}_{B}(p_1,p_2,p_3,p_4)\\
&&+\mathcal{M}^{\mu}_{C}(p_1,p_2,p_3,p_4)+\mathcal{M}^{\mu}_{C}(p_3,p_4,p_1,p_2).\nonumber
\end{eqnarray}

The final gauge invariant amplitude for this set of diagrams is given by Eq. (\ref{abc24}).
A similar expression is obtained for the exchange $p_2 \leftrightarrow p_4$, by adding the proper amplitudes Eq.~(\ref{abc42}) and replacing $Q_{1} \equiv s_{34}-s_{12}\ \text{with} \ Q_{2} \equiv s_{23}-s_{41}$.

\begin{acknowledgments}
We acknowledge the support of SECIHTI, Mexico Grant No. 520992 (LAJP) and DGAPA-PAPIIT UNAM, under Grant No. IN105526
\end{acknowledgments}

\bibliography{Ksmdm}

\begin{thebibliography}{62}%
\makeatletter
\providecommand \@ifxundefined [1]{%
 \@ifx{#1\undefined}
}%
\providecommand \@ifnum [1]{%
 \ifnum #1\expandafter \@firstoftwo
 \else \expandafter \@secondoftwo
 \fi
}%
\providecommand \@ifx [1]{%
 \ifx #1\expandafter \@firstoftwo
 \else \expandafter \@secondoftwo
 \fi
}%
\providecommand \natexlab [1]{#1}%
\providecommand \enquote  [1]{``#1''}%
\providecommand \bibnamefont  [1]{#1}%
\providecommand \bibfnamefont [1]{#1}%
\providecommand \citenamefont [1]{#1}%
\providecommand \href@noop [0]{\@secondoftwo}%
\providecommand \href [0]{\begingroup \@sanitize@url \@href}%
\providecommand \@href[1]{\@@startlink{#1}\@@href}%
\providecommand \@@href[1]{\endgroup#1\@@endlink}%
\providecommand \@sanitize@url [0]{\catcode `\\12\catcode `\$12\catcode `\&12\catcode `\#12\catcode `\^12\catcode `\_12\catcode `\%12\relax}%
\providecommand \@@startlink[1]{}%
\providecommand \@@endlink[0]{}%
\providecommand \url  [0]{\begingroup\@sanitize@url \@url }%
\providecommand \@url [1]{\endgroup\@href {#1}{\urlprefix }}%
\providecommand \urlprefix  [0]{URL }%
\providecommand \Eprint [0]{\href }%
\providecommand \doibase [0]{https://doi.org/}%
\providecommand \selectlanguage [0]{\@gobble}%
\providecommand \bibinfo  [0]{\@secondoftwo}%
\providecommand \bibfield  [0]{\@secondoftwo}%
\providecommand \translation [1]{[#1]}%
\providecommand \BibitemOpen [0]{}%
\providecommand \bibitemStop [0]{}%
\providecommand \bibitemNoStop [0]{.\EOS\space}%
\providecommand \EOS [0]{\spacefactor3000\relax}%
\providecommand \BibitemShut  [1]{\csname bibitem#1\endcsname}%
\let\auto@bib@innerbib\@empty
\bibitem [{\citenamefont {Deser}\ \emph {et~al.}(1970)\citenamefont {Deser}, \citenamefont {Grisaru},\ and\ \citenamefont {Pendleton}}]{Deser:1970spa}%
  \BibitemOpen
  \bibinfo {editor} {\bibfnamefont {S.~D.}\ \bibnamefont {Deser}}, \bibinfo {editor} {\bibfnamefont {M.~T.}\ \bibnamefont {Grisaru}},\ and\ \bibinfo {editor} {\bibfnamefont {H.}~\bibnamefont {Pendleton}},\ eds.,\ \href@noop {} {\emph {\bibinfo {title} {{Proceedings, 13th Brandeis University Summer Institute in Theoretical Physics, Lectures On Elementary Particles and Quantum Field Theory}: {Waltham, MA, USA, June 15 - July 24 1970}}}}\ (\bibinfo  {publisher} {MIT},\ \bibinfo {address} {Cambridge, MA, USA},\ \bibinfo {year} {1970})\BibitemShut {NoStop}%
\bibitem [{\citenamefont {Brodsky}\ and\ \citenamefont {Hiller}(1992)}]{Brodsky:1992px}%
  \BibitemOpen
  \bibfield  {author} {\bibinfo {author} {\bibfnamefont {S.~J.}\ \bibnamefont {Brodsky}}\ and\ \bibinfo {author} {\bibfnamefont {J.~R.}\ \bibnamefont {Hiller}},\ }\bibfield  {title} {\bibinfo {title} {{Universal properties of the electromagnetic interactions of spin one systems}},\ }\href {https://doi.org/10.1103/PhysRevD.46.2141} {\bibfield  {journal} {\bibinfo  {journal} {Phys. Rev. D}\ }\textbf {\bibinfo {volume} {46}},\ \bibinfo {pages} {2141} (\bibinfo {year} {1992})}\BibitemShut {NoStop}%
\bibitem [{\citenamefont {Ferrara}\ \emph {et~al.}(1992)\citenamefont {Ferrara}, \citenamefont {Porrati},\ and\ \citenamefont {Telegdi}}]{Ferrara:1992yc}%
  \BibitemOpen
  \bibfield  {author} {\bibinfo {author} {\bibfnamefont {S.}~\bibnamefont {Ferrara}}, \bibinfo {author} {\bibfnamefont {M.}~\bibnamefont {Porrati}},\ and\ \bibinfo {author} {\bibfnamefont {V.~L.}\ \bibnamefont {Telegdi}},\ }\bibfield  {title} {\bibinfo {title} {{$g=2$ as the natural value of the tree-level gyromagnetic ratio of elementary particles}},\ }\href {https://doi.org/10.1103/PhysRevD.46.3529} {\bibfield  {journal} {\bibinfo  {journal} {Phys. Rev. D}\ }\textbf {\bibinfo {volume} {46}},\ \bibinfo {pages} {3529} (\bibinfo {year} {1992})}\BibitemShut {NoStop}%
\bibitem [{\citenamefont {Jackiw}(1998)}]{Jackiw:1997vw}%
  \BibitemOpen
  \bibfield  {author} {\bibinfo {author} {\bibfnamefont {R.}~\bibnamefont {Jackiw}},\ }\bibfield  {title} {\bibinfo {title} {{G = 2 as a gauge condition}},\ }\href {https://doi.org/10.1103/PhysRevD.57.2635} {\bibfield  {journal} {\bibinfo  {journal} {Phys. Rev. D}\ }\textbf {\bibinfo {volume} {57}},\ \bibinfo {pages} {2635} (\bibinfo {year} {1998})}\BibitemShut {NoStop}%
\bibitem [{\citenamefont {Toledo~Sanchez}(2002)}]{ToledoSanchez:2002rn}%
  \BibitemOpen
  \bibfield  {author} {\bibinfo {author} {\bibfnamefont {G.}~\bibnamefont {Toledo~Sanchez}},\ }\bibfield  {title} {\bibinfo {title} {{Structure of radiative interferences and g=2 for vector mesons}},\ }\href {https://doi.org/10.1103/PhysRevD.66.097301} {\bibfield  {journal} {\bibinfo  {journal} {Phys. Rev. D}\ }\textbf {\bibinfo {volume} {66}},\ \bibinfo {pages} {097301} (\bibinfo {year} {2002})}\BibitemShut {NoStop}%
\bibitem [{\citenamefont {Holstein}(2006)}]{Holstein:2006wi}%
  \BibitemOpen
  \bibfield  {author} {\bibinfo {author} {\bibfnamefont {B.~R.}\ \bibnamefont {Holstein}},\ }\bibfield  {title} {\bibinfo {title} {{How Large is the 'natural' magnetic moment?}},\ }\href {https://doi.org/10.1119/1.2345655} {\bibfield  {journal} {\bibinfo  {journal} {Am. J. Phys.}\ }\textbf {\bibinfo {volume} {74}},\ \bibinfo {pages} {1104} (\bibinfo {year} {2006})}\BibitemShut {NoStop}%
\bibitem [{\citenamefont {Dbeyssi}\ \emph {et~al.}(2012)\citenamefont {Dbeyssi}, \citenamefont {Tomasi-Gustafsson}, \citenamefont {Gakh},\ and\ \citenamefont {Adamuscin}}]{Dbeyssi:2011ep}%
  \BibitemOpen
  \bibfield  {author} {\bibinfo {author} {\bibfnamefont {A.}~\bibnamefont {Dbeyssi}}, \bibinfo {author} {\bibfnamefont {E.}~\bibnamefont {Tomasi-Gustafsson}}, \bibinfo {author} {\bibfnamefont {G.~I.}\ \bibnamefont {Gakh}},\ and\ \bibinfo {author} {\bibfnamefont {C.}~\bibnamefont {Adamuscin}},\ }\bibfield  {title} {\bibinfo {title} {{Experimental constraint on the $\rho -$ meson form factors in the time--like region}},\ }\href {https://doi.org/10.1103/PhysRevC.85.048201} {\bibfield  {journal} {\bibinfo  {journal} {Phys. Rev. C}\ }\textbf {\bibinfo {volume} {85}},\ \bibinfo {pages} {048201} (\bibinfo {year} {2012})}\BibitemShut {NoStop}%
\bibitem [{\citenamefont {Lorce}(2009)}]{Lorce:2009bs}%
  \BibitemOpen
  \bibfield  {author} {\bibinfo {author} {\bibfnamefont {C.}~\bibnamefont {Lorce}},\ }\bibfield  {title} {\bibinfo {title} {{Electromagnetic properties for arbitrary spin particles: Natural electromagnetic moments from light-cone arguments}},\ }\href {https://doi.org/10.1103/PhysRevD.79.113011} {\bibfield  {journal} {\bibinfo  {journal} {Phys. Rev. D}\ }\textbf {\bibinfo {volume} {79}},\ \bibinfo {pages} {113011} (\bibinfo {year} {2009})}\BibitemShut {NoStop}%
\bibitem [{\citenamefont {Delgado-Acosta}\ \emph {et~al.}(2012)\citenamefont {Delgado-Acosta}, \citenamefont {Kirchbach}, \citenamefont {Napsuciale},\ and\ \citenamefont {Rodriguez}}]{Delgado-Acosta:2012dxv}%
  \BibitemOpen
  \bibfield  {author} {\bibinfo {author} {\bibfnamefont {E.~G.}\ \bibnamefont {Delgado-Acosta}}, \bibinfo {author} {\bibfnamefont {M.}~\bibnamefont {Kirchbach}}, \bibinfo {author} {\bibfnamefont {M.}~\bibnamefont {Napsuciale}},\ and\ \bibinfo {author} {\bibfnamefont {S.}~\bibnamefont {Rodriguez}},\ }\bibfield  {title} {\bibinfo {title} {{Electromagnetic multipole moments of elementary spin-1/2, 1, and 3/2 particles}},\ }\href {https://doi.org/10.1103/PhysRevD.85.116006} {\bibfield  {journal} {\bibinfo  {journal} {Phys. Rev. D}\ }\textbf {\bibinfo {volume} {85}},\ \bibinfo {pages} {116006} (\bibinfo {year} {2012})}\BibitemShut {NoStop}%
\bibitem [{\citenamefont {Haberzettl}(2019)}]{Haberzettl:2019qpa}%
  \BibitemOpen
  \bibfield  {author} {\bibinfo {author} {\bibfnamefont {H.}~\bibnamefont {Haberzettl}},\ }\bibfield  {title} {\bibinfo {title} {{Model-independent form-factor constraints for electromagnetic spin-1 currents}},\ }\href {https://doi.org/10.1103/PhysRevD.100.036008} {\bibfield  {journal} {\bibinfo  {journal} {Phys. Rev. D}\ }\textbf {\bibinfo {volume} {100}},\ \bibinfo {pages} {036008} (\bibinfo {year} {2019})}\BibitemShut {NoStop}%
\bibitem [{\citenamefont {Bagdasaryan}\ \emph {et~al.}(1985)\citenamefont {Bagdasaryan}, \citenamefont {Esaibegian},\ and\ \citenamefont {Ter-Isaakian}}]{Bagdasaryan:1984kz}%
  \BibitemOpen
  \bibfield  {author} {\bibinfo {author} {\bibfnamefont {A.~S.}\ \bibnamefont {Bagdasaryan}}, \bibinfo {author} {\bibfnamefont {S.~V.}\ \bibnamefont {Esaibegian}},\ and\ \bibinfo {author} {\bibfnamefont {N.~L.}\ \bibnamefont {Ter-Isaakian}},\ }\bibfield  {title} {\bibinfo {title} {Form-factors of mesons and meson resonances at small and intermediate momentum transfers $q^2$ in the relativistic quark model},\ }\href {https://inspirehep.net/literature/210027} {\bibfield  {journal} {\bibinfo  {journal} {Yad. Fiz.}\ }\textbf {\bibinfo {volume} {42}},\ \bibinfo {pages} {440} (\bibinfo {year} {1985})}\BibitemShut {NoStop}%
\bibitem [{\citenamefont {Cardarelli}\ \emph {et~al.}(1995)\citenamefont {Cardarelli}, \citenamefont {Grach}, \citenamefont {Narodetsky}, \citenamefont {Salme},\ and\ \citenamefont {Simula}}]{Cardarelli:1994yq}%
  \BibitemOpen
  \bibfield  {author} {\bibinfo {author} {\bibfnamefont {F.}~\bibnamefont {Cardarelli}}, \bibinfo {author} {\bibfnamefont {I.~L.}\ \bibnamefont {Grach}}, \bibinfo {author} {\bibfnamefont {I.~M.}\ \bibnamefont {Narodetsky}}, \bibinfo {author} {\bibfnamefont {G.}~\bibnamefont {Salme}},\ and\ \bibinfo {author} {\bibfnamefont {S.}~\bibnamefont {Simula}},\ }\bibfield  {title} {\bibinfo {title} {{Electromagnetic form-factors of the rho meson in a light front constituent quark model}},\ }\href {https://doi.org/10.1016/0370-2693(95)00230-I} {\bibfield  {journal} {\bibinfo  {journal} {Phys. Lett. B}\ }\textbf {\bibinfo {volume} {349}},\ \bibinfo {pages} {393} (\bibinfo {year} {1995})}\BibitemShut {NoStop}%
\bibitem [{\citenamefont {de~Melo}\ and\ \citenamefont {Frederico}(1997)}]{deMelo:1997hh}%
  \BibitemOpen
  \bibfield  {author} {\bibinfo {author} {\bibfnamefont {J.~P. B.~C.}\ \bibnamefont {de~Melo}}\ and\ \bibinfo {author} {\bibfnamefont {T.}~\bibnamefont {Frederico}},\ }\bibfield  {title} {\bibinfo {title} {{Covariant and light front approaches to the rho meson electromagnetic form-factors}},\ }\href {https://doi.org/10.1103/PhysRevC.55.2043} {\bibfield  {journal} {\bibinfo  {journal} {Phys. Rev. C}\ }\textbf {\bibinfo {volume} {55}},\ \bibinfo {pages} {2043} (\bibinfo {year} {1997})}\BibitemShut {NoStop}%
\bibitem [{\citenamefont {Melikhov}\ and\ \citenamefont {Simula}(2002)}]{Melikhov:2001pm}%
  \BibitemOpen
  \bibfield  {author} {\bibinfo {author} {\bibfnamefont {D.}~\bibnamefont {Melikhov}}\ and\ \bibinfo {author} {\bibfnamefont {S.}~\bibnamefont {Simula}},\ }\bibfield  {title} {\bibinfo {title} {{Electromagnetic form-factors in the light front formalism and the Feynman triangle diagram: Spin 0 and spin 1 two fermion systems}},\ }\href {https://doi.org/10.1103/PhysRevD.65.094043} {\bibfield  {journal} {\bibinfo  {journal} {Phys. Rev. D}\ }\textbf {\bibinfo {volume} {65}},\ \bibinfo {pages} {094043} (\bibinfo {year} {2002})}\BibitemShut {NoStop}%
\bibitem [{\citenamefont {Samsonov}(2003)}]{Alexander_2003}%
  \BibitemOpen
  \bibfield  {author} {\bibinfo {author} {\bibfnamefont {A.}~\bibnamefont {Samsonov}},\ }\bibfield  {title} {\bibinfo {title} {{Magnetic moment of the rho meson in QCD sum rules: alpha(s) corrections}},\ }\href {https://doi.org/10.1088/1126-6708/2003/12/061} {\bibfield  {journal} {\bibinfo  {journal} {JHEP}\ }\textbf {\bibinfo {volume} {12}},\ \bibinfo {pages} {061}}\BibitemShut {NoStop}%
\bibitem [{\citenamefont {Aliev}\ \emph {et~al.}(2003)\citenamefont {Aliev}, \citenamefont {Kanik},\ and\ \citenamefont {Savci}}]{Aliev:2003ba}%
  \BibitemOpen
  \bibfield  {author} {\bibinfo {author} {\bibfnamefont {T.~M.}\ \bibnamefont {Aliev}}, \bibinfo {author} {\bibfnamefont {I.}~\bibnamefont {Kanik}},\ and\ \bibinfo {author} {\bibfnamefont {M.}~\bibnamefont {Savci}},\ }\bibfield  {title} {\bibinfo {title} {{magnetic moment of the rho meson in qcd light cone sum rules}},\ }\href {https://doi.org/10.1103/PhysRevD.68.056002} {\bibfield  {journal} {\bibinfo  {journal} {Phys. Rev. D}\ }\textbf {\bibinfo {volume} {68}},\ \bibinfo {pages} {056002} (\bibinfo {year} {2003})}\BibitemShut {NoStop}%
\bibitem [{\citenamefont {Jaus}(2003)}]{Jaus:2002sv}%
  \BibitemOpen
  \bibfield  {author} {\bibinfo {author} {\bibfnamefont {W.}~\bibnamefont {Jaus}},\ }\bibfield  {title} {\bibinfo {title} {{Consistent treatment of spin 1 mesons in the light front quark model}},\ }\href {https://doi.org/10.1103/PhysRevD.67.094010} {\bibfield  {journal} {\bibinfo  {journal} {Phys. Rev. D}\ }\textbf {\bibinfo {volume} {67}},\ \bibinfo {pages} {094010} (\bibinfo {year} {2003})}\BibitemShut {NoStop}%
\bibitem [{\citenamefont {Choi}\ and\ \citenamefont {Ji}(2004)}]{Choi:2004ww}%
  \BibitemOpen
  \bibfield  {author} {\bibinfo {author} {\bibfnamefont {H.-M.}\ \bibnamefont {Choi}}\ and\ \bibinfo {author} {\bibfnamefont {C.-R.}\ \bibnamefont {Ji}},\ }\bibfield  {title} {\bibinfo {title} {{Electromagnetic structure of the rho meson in the light front quark model}},\ }\href {https://doi.org/10.1103/PhysRevD.70.053015} {\bibfield  {journal} {\bibinfo  {journal} {Phys. Rev. D}\ }\textbf {\bibinfo {volume} {70}},\ \bibinfo {pages} {053015} (\bibinfo {year} {2004})}\BibitemShut {NoStop}%
\bibitem [{\citenamefont {He}\ \emph {et~al.}(2004)\citenamefont {He}, \citenamefont {Julia-Diaz},\ and\ \citenamefont {Dong}}]{He:2004ba}%
  \BibitemOpen
  \bibfield  {author} {\bibinfo {author} {\bibfnamefont {J.}~\bibnamefont {He}}, \bibinfo {author} {\bibfnamefont {B.}~\bibnamefont {Julia-Diaz}},\ and\ \bibinfo {author} {\bibfnamefont {Y.-b.}\ \bibnamefont {Dong}},\ }\bibfield  {title} {\bibinfo {title} {{Electromagnetic form-factors of pion and rho in the three forms of relativistic kinematics}},\ }\href {https://doi.org/10.1016/j.physletb.2004.10.004} {\bibfield  {journal} {\bibinfo  {journal} {Phys. Lett. B}\ }\textbf {\bibinfo {volume} {602}},\ \bibinfo {pages} {212} (\bibinfo {year} {2004})}\BibitemShut {NoStop}%
\bibitem [{\citenamefont {Bhagwat}\ \emph {et~al.}(2009)\citenamefont {Bhagwat}, \citenamefont {Özpineci},\ and\ \citenamefont {Savcı}}]{ALIEV2009470}%
  \BibitemOpen
  \bibfield  {author} {\bibinfo {author} {\bibfnamefont {T.}~\bibnamefont {Bhagwat}}, \bibinfo {author} {\bibfnamefont {A.}~\bibnamefont {Özpineci}},\ and\ \bibinfo {author} {\bibfnamefont {M.}~\bibnamefont {Savcı}},\ }\bibfield  {title} {\bibinfo {title} {Magnetic and quadrupole moments of light spin-1 mesons in light cone qcd sum rules},\ }\href {https://doi.org/https://doi.org/10.1016/j.physletb.2009.06.073} {\bibfield  {journal} {\bibinfo  {journal} {Physics Letters B}\ }\textbf {\bibinfo {volume} {678}},\ \bibinfo {pages} {470} (\bibinfo {year} {2009})}\BibitemShut {NoStop}%
\bibitem [{\citenamefont {Biernat}\ and\ \citenamefont {Schweiger}(2014)}]{Biernat:2014dea}%
  \BibitemOpen
  \bibfield  {author} {\bibinfo {author} {\bibfnamefont {E.~P.}\ \bibnamefont {Biernat}}\ and\ \bibinfo {author} {\bibfnamefont {W.}~\bibnamefont {Schweiger}},\ }\bibfield  {title} {\bibinfo {title} {{Electromagnetic rho-meson form factors in point-form relativistic quantum mechanics}},\ }\href {https://doi.org/10.1103/PhysRevC.89.055205} {\bibfield  {journal} {\bibinfo  {journal} {Phys. Rev. C}\ }\textbf {\bibinfo {volume} {89}},\ \bibinfo {pages} {055205} (\bibinfo {year} {2014})}\BibitemShut {NoStop}%
\bibitem [{\citenamefont {Simonis}(2016)}]{Simonis2016}%
  \BibitemOpen
  \bibfield  {author} {\bibinfo {author} {\bibfnamefont {V.}~\bibnamefont {Simonis}},\ }\bibfield  {title} {\bibinfo {title} {Magnetic properties of ground-state mesons.},\ }\href {https://doi.org/10.1140/epja/i2016-16090-5} {\bibfield  {journal} {\bibinfo  {journal} {The European Physical Journal A}\ }\textbf {\bibinfo {volume} {52}},\ \bibinfo {pages} {90} (\bibinfo {year} {2016})}\BibitemShut {NoStop}%
\bibitem [{\citenamefont {de~Melo}\ and\ \citenamefont {Tsushima}(2017)}]{deMelo:2016ynt}%
  \BibitemOpen
  \bibfield  {author} {\bibinfo {author} {\bibfnamefont {J.~P. B.~C.}\ \bibnamefont {de~Melo}}\ and\ \bibinfo {author} {\bibfnamefont {K.}~\bibnamefont {Tsushima}},\ }\bibfield  {title} {\bibinfo {title} {{In-Medium $\rho$-Meson Properties in a Light-Front Approach}},\ }\href {https://doi.org/10.1007/s00601-017-1233-2} {\bibfield  {journal} {\bibinfo  {journal} {Few Body Syst.}\ }\textbf {\bibinfo {volume} {58}},\ \bibinfo {pages} {82} (\bibinfo {year} {2017})}\BibitemShut {NoStop}%
\bibitem [{\citenamefont {Krutov}\ \emph {et~al.}(2018)\citenamefont {Krutov}, \citenamefont {Polezhaev},\ and\ \citenamefont {Troitsky}}]{Krutov:2018mbu}%
  \BibitemOpen
  \bibfield  {author} {\bibinfo {author} {\bibfnamefont {A.~F.}\ \bibnamefont {Krutov}}, \bibinfo {author} {\bibfnamefont {R.~G.}\ \bibnamefont {Polezhaev}},\ and\ \bibinfo {author} {\bibfnamefont {V.~E.}\ \bibnamefont {Troitsky}},\ }\bibfield  {title} {\bibinfo {title} {{Magnetic moment of the \ensuremath{\rho} meson in instant-form relativistic quantum mechanics}},\ }\href {https://doi.org/10.1103/PhysRevD.97.033007} {\bibfield  {journal} {\bibinfo  {journal} {Phys. Rev. D}\ }\textbf {\bibinfo {volume} {97}},\ \bibinfo {pages} {033007} (\bibinfo {year} {2018})}\BibitemShut {NoStop}%
\bibitem [{\citenamefont {Simonis}(2018)}]{Simonis:2018rld}%
  \BibitemOpen
  \bibfield  {author} {\bibinfo {author} {\bibfnamefont {V.}~\bibnamefont {Simonis}},\ }\href@noop {} {\bibinfo {title} {Improved predictions for magnetic moments and m1 decay widths of heavy hadrons}} (\bibinfo {year} {2018}),\ \Eprint {https://arxiv.org/abs/1803.01809} {arXiv:1803.01809 [hep-ph]} \BibitemShut {NoStop}%
\bibitem [{\citenamefont {De~Melo}(2019)}]{DeMelo:2018bim}%
  \BibitemOpen
  \bibfield  {author} {\bibinfo {author} {\bibfnamefont {J.~P. B.~C.}\ \bibnamefont {De~Melo}},\ }\bibfield  {title} {\bibinfo {title} {{Unambiguous Extraction of the Electromagnetic Form Factors for Spin-1 Particles on the Light-Front}},\ }\href {https://doi.org/10.1016/j.physletb.2018.11.003} {\bibfield  {journal} {\bibinfo  {journal} {Phys. Lett. B}\ }\textbf {\bibinfo {volume} {788}},\ \bibinfo {pages} {152} (\bibinfo {year} {2019})}\BibitemShut {NoStop}%
\bibitem [{\citenamefont {Hawes}\ and\ \citenamefont {Pichowsky}(1999)}]{Hawes:1998bz}%
  \BibitemOpen
  \bibfield  {author} {\bibinfo {author} {\bibfnamefont {F.~T.}\ \bibnamefont {Hawes}}\ and\ \bibinfo {author} {\bibfnamefont {M.~A.}\ \bibnamefont {Pichowsky}},\ }\bibfield  {title} {\bibinfo {title} {{Electromagnetic form-factors of light vector mesons}},\ }\href {https://doi.org/10.1103/PhysRevC.59.1743} {\bibfield  {journal} {\bibinfo  {journal} {Phys. Rev. C}\ }\textbf {\bibinfo {volume} {59}},\ \bibinfo {pages} {1743} (\bibinfo {year} {1999})}\BibitemShut {NoStop}%
\bibitem [{\citenamefont {Bhagwat}\ and\ \citenamefont {Maris}(2008)}]{Bhagwat:2006pu}%
  \BibitemOpen
  \bibfield  {author} {\bibinfo {author} {\bibfnamefont {M.~S.}\ \bibnamefont {Bhagwat}}\ and\ \bibinfo {author} {\bibfnamefont {P.}~\bibnamefont {Maris}},\ }\bibfield  {title} {\bibinfo {title} {{Vector meson form factors and their quark-mass dependence}},\ }\href {https://doi.org/10.1103/PhysRevC.77.025203} {\bibfield  {journal} {\bibinfo  {journal} {Phys. Rev. C}\ }\textbf {\bibinfo {volume} {77}},\ \bibinfo {pages} {025203} (\bibinfo {year} {2008})}\BibitemShut {NoStop}%
\bibitem [{\citenamefont {Roberts}\ \emph {et~al.}(2011)\citenamefont {Roberts}, \citenamefont {Bashir}, \citenamefont {Gutierrez-Guerrero}, \citenamefont {Roberts},\ and\ \citenamefont {Wilson}}]{Roberts:2011wy}%
  \BibitemOpen
  \bibfield  {author} {\bibinfo {author} {\bibfnamefont {H.~L.~L.}\ \bibnamefont {Roberts}}, \bibinfo {author} {\bibfnamefont {A.}~\bibnamefont {Bashir}}, \bibinfo {author} {\bibfnamefont {L.~X.}\ \bibnamefont {Gutierrez-Guerrero}}, \bibinfo {author} {\bibfnamefont {C.~D.}\ \bibnamefont {Roberts}},\ and\ \bibinfo {author} {\bibfnamefont {D.~J.}\ \bibnamefont {Wilson}},\ }\bibfield  {title} {\bibinfo {title} {{pi- and rho-mesons, and their diquark partners, from a contact interaction}},\ }\href {https://doi.org/10.1103/PhysRevC.83.065206} {\bibfield  {journal} {\bibinfo  {journal} {Phys. Rev. C}\ }\textbf {\bibinfo {volume} {83}},\ \bibinfo {pages} {065206} (\bibinfo {year} {2011})}\BibitemShut {NoStop}%
\bibitem [{\citenamefont {Pitschmann}\ \emph {et~al.}(2013)\citenamefont {Pitschmann}, \citenamefont {Seng}, \citenamefont {Ramsey-Musolf}, \citenamefont {Roberts}, \citenamefont {Schmidt},\ and\ \citenamefont {Wilson}}]{Pitschmann:2012by}%
  \BibitemOpen
  \bibfield  {author} {\bibinfo {author} {\bibfnamefont {M.}~\bibnamefont {Pitschmann}}, \bibinfo {author} {\bibfnamefont {C.-Y.}\ \bibnamefont {Seng}}, \bibinfo {author} {\bibfnamefont {M.~J.}\ \bibnamefont {Ramsey-Musolf}}, \bibinfo {author} {\bibfnamefont {C.~D.}\ \bibnamefont {Roberts}}, \bibinfo {author} {\bibfnamefont {S.~M.}\ \bibnamefont {Schmidt}},\ and\ \bibinfo {author} {\bibfnamefont {D.~J.}\ \bibnamefont {Wilson}},\ }\bibfield  {title} {\bibinfo {title} {{Electric dipole moment of the \ensuremath{\rho} meson}},\ }\href {https://doi.org/10.1103/PhysRevC.87.015205} {\bibfield  {journal} {\bibinfo  {journal} {Phys. Rev. C}\ }\textbf {\bibinfo {volume} {87}},\ \bibinfo {pages} {015205} (\bibinfo {year} {2013})}\BibitemShut {NoStop}%
\bibitem [{\citenamefont {Xu}\ \emph {et~al.}(2019)\citenamefont {Xu}, \citenamefont {Binosi}, \citenamefont {Cui}, \citenamefont {Li}, \citenamefont {Roberts}, \citenamefont {Xu},\ and\ \citenamefont {Zong}}]{Xu:2019ilh}%
  \BibitemOpen
  \bibfield  {author} {\bibinfo {author} {\bibfnamefont {Y.-Z.}\ \bibnamefont {Xu}}, \bibinfo {author} {\bibfnamefont {D.}~\bibnamefont {Binosi}}, \bibinfo {author} {\bibfnamefont {Z.-F.}\ \bibnamefont {Cui}}, \bibinfo {author} {\bibfnamefont {B.-L.}\ \bibnamefont {Li}}, \bibinfo {author} {\bibfnamefont {C.~D.}\ \bibnamefont {Roberts}}, \bibinfo {author} {\bibfnamefont {S.-S.}\ \bibnamefont {Xu}},\ and\ \bibinfo {author} {\bibfnamefont {H.~S.}\ \bibnamefont {Zong}},\ }\bibfield  {title} {\bibinfo {title} {{Elastic electromagnetic form factors of vector mesons}},\ }\href {https://doi.org/10.1103/PhysRevD.100.114038} {\bibfield  {journal} {\bibinfo  {journal} {Phys. Rev. D}\ }\textbf {\bibinfo {volume} {100}},\ \bibinfo {pages} {114038} (\bibinfo {year} {2019})}\BibitemShut {NoStop}%
\bibitem [{\citenamefont {Xing}\ \emph {et~al.}(2021)\citenamefont {Xing}, \citenamefont {Raya},\ and\ \citenamefont {Chang}}]{Xing:2021dwe}%
  \BibitemOpen
  \bibfield  {author} {\bibinfo {author} {\bibfnamefont {Z.}~\bibnamefont {Xing}}, \bibinfo {author} {\bibfnamefont {K.}~\bibnamefont {Raya}},\ and\ \bibinfo {author} {\bibfnamefont {L.}~\bibnamefont {Chang}},\ }\bibfield  {title} {\bibinfo {title} {{Quark anomalous magnetic moment and its effects on the \ensuremath{\rho} meson properties}},\ }\href {https://doi.org/10.1103/PhysRevD.104.054038} {\bibfield  {journal} {\bibinfo  {journal} {Phys. Rev. D}\ }\textbf {\bibinfo {volume} {104}},\ \bibinfo {pages} {054038} (\bibinfo {year} {2021})}\BibitemShut {NoStop}%
\bibitem [{\citenamefont {Shi}\ \emph {et~al.}(2023)\citenamefont {Shi}, \citenamefont {Li}, \citenamefont {Yin},\ and\ \citenamefont {Jia}}]{Shi:2023oll}%
  \BibitemOpen
  \bibfield  {author} {\bibinfo {author} {\bibfnamefont {C.}~\bibnamefont {Shi}}, \bibinfo {author} {\bibfnamefont {J.}~\bibnamefont {Li}}, \bibinfo {author} {\bibfnamefont {P.-L.}\ \bibnamefont {Yin}},\ and\ \bibinfo {author} {\bibfnamefont {W.}~\bibnamefont {Jia}},\ }\bibfield  {title} {\bibinfo {title} {{Unpolarized generalized parton distributions of light and heavy vector mesons}},\ }\href {https://doi.org/10.1103/PhysRevD.107.074009} {\bibfield  {journal} {\bibinfo  {journal} {Phys. Rev. D}\ }\textbf {\bibinfo {volume} {107}},\ \bibinfo {pages} {074009} (\bibinfo {year} {2023})}\BibitemShut {NoStop}%
\bibitem [{\citenamefont {Xu}\ and\ \citenamefont {Segovia}(2023)}]{Xu:2023vlt}%
  \BibitemOpen
  \bibfield  {author} {\bibinfo {author} {\bibfnamefont {Y.-Z.}\ \bibnamefont {Xu}}\ and\ \bibinfo {author} {\bibfnamefont {J.}~\bibnamefont {Segovia}},\ }\bibfield  {title} {\bibinfo {title} {{An Assessment of Pseudoscalar and Vector Meson Electromagnetic Form Factors}},\ }\href {https://doi.org/10.1007/s00601-023-01845-6} {\bibfield  {journal} {\bibinfo  {journal} {Few Body Syst.}\ }\textbf {\bibinfo {volume} {64}},\ \bibinfo {pages} {62} (\bibinfo {year} {2023})}\BibitemShut {NoStop}%
\bibitem [{\citenamefont {Haurysh}\ and\ \citenamefont {Andreev}(2021)}]{Haurysh2021}%
  \BibitemOpen
  \bibfield  {author} {\bibinfo {author} {\bibfnamefont {V.~Y.}\ \bibnamefont {Haurysh}}\ and\ \bibinfo {author} {\bibfnamefont {V.~V.}\ \bibnamefont {Andreev}},\ }\bibfield  {title} {\bibinfo {title} {$\rho$ -meson form-factors in point form of poincaré-invariant quantum mechanics},\ }\href {https://doi.org/10.1007/s00601-021-01610-7} {\bibfield  {journal} {\bibinfo  {journal} {Few-Body Systems}\ ,\ \bibinfo {pages} {29}} (\bibinfo {year} {2021})}\BibitemShut {NoStop}%
\bibitem [{\citenamefont {Zhang}\ \emph {et~al.}(2022)\citenamefont {Zhang}, \citenamefont {Kang},\ and\ \citenamefont {Ping}}]{Zhang:2022zim}%
  \BibitemOpen
  \bibfield  {author} {\bibinfo {author} {\bibfnamefont {J.-L.}\ \bibnamefont {Zhang}}, \bibinfo {author} {\bibfnamefont {G.-Z.}\ \bibnamefont {Kang}},\ and\ \bibinfo {author} {\bibfnamefont {J.-L.}\ \bibnamefont {Ping}},\ }\bibfield  {title} {\bibinfo {title} {{\ensuremath{\rho} meson generalized parton distributions in the Nambu\textendash{}Jona-Lasinio model}},\ }\href {https://doi.org/10.1103/PhysRevD.105.094015} {\bibfield  {journal} {\bibinfo  {journal} {Phys. Rev. D}\ }\textbf {\bibinfo {volume} {105}},\ \bibinfo {pages} {094015} (\bibinfo {year} {2022})}\BibitemShut {NoStop}%
\bibitem [{\citenamefont {Luan}\ \emph {et~al.}(2015)\citenamefont {Luan}, \citenamefont {Chen},\ and\ \citenamefont {Deng}}]{Luan_2015}%
  \BibitemOpen
  \bibfield  {author} {\bibinfo {author} {\bibfnamefont {Y.-L.}\ \bibnamefont {Luan}}, \bibinfo {author} {\bibfnamefont {X.-L.}\ \bibnamefont {Chen}},\ and\ \bibinfo {author} {\bibfnamefont {W.-Z.}\ \bibnamefont {Deng}},\ }\bibfield  {title} {\bibinfo {title} {Meson electro-magnetic form factors in an extended nambu–jona-lasinio model including heavy quark flavors},\ }\href {https://doi.org/10.1088/1674-1137/39/11/113103} {\bibfield  {journal} {\bibinfo  {journal} {Chinese Physics C}\ }\textbf {\bibinfo {volume} {39}},\ \bibinfo {pages} {113103} (\bibinfo {year} {2015})}\BibitemShut {NoStop}%
\bibitem [{\citenamefont {Djukanovic}\ \emph {et~al.}(2014)\citenamefont {Djukanovic}, \citenamefont {Epelbaum}, \citenamefont {Gegelia},\ and\ \citenamefont {Meißner}}]{Djukanovic:2013mka}%
  \BibitemOpen
  \bibfield  {author} {\bibinfo {author} {\bibfnamefont {D.}~\bibnamefont {Djukanovic}}, \bibinfo {author} {\bibfnamefont {E.}~\bibnamefont {Epelbaum}}, \bibinfo {author} {\bibfnamefont {J.}~\bibnamefont {Gegelia}},\ and\ \bibinfo {author} {\bibfnamefont {U.-G.}\ \bibnamefont {Meißner}},\ }\bibfield  {title} {\bibinfo {title} {{The magnetic moment of the $\rho$-meson}},\ }\href {https://doi.org/10.1016/j.physletb.2014.01.001} {\bibfield  {journal} {\bibinfo  {journal} {Phys. Lett. B}\ }\textbf {\bibinfo {volume} {730}},\ \bibinfo {pages} {115} (\bibinfo {year} {2014})}\BibitemShut {NoStop}%
\bibitem [{\citenamefont {Rojas}\ and\ \citenamefont {Toledo}(2024)}]{Rojas:2024tmn}%
  \BibitemOpen
  \bibfield  {author} {\bibinfo {author} {\bibfnamefont {A.}~\bibnamefont {Rojas}}\ and\ \bibinfo {author} {\bibfnamefont {G.}~\bibnamefont {Toledo}},\ }\bibfield  {title} {\bibinfo {title} {{{\ensuremath{\rho}} meson magnetic dipole moment from BaBar e+e-{\textrightarrow}{\ensuremath{\pi}}+{\ensuremath{\pi}}-{\ensuremath{\pi}}0{\ensuremath{\pi}}0 cross section measurement}},\ }\href {https://doi.org/10.1103/PhysRevD.110.056037} {\bibfield  {journal} {\bibinfo  {journal} {Phys. Rev. D}\ }\textbf {\bibinfo {volume} {110}},\ \bibinfo {pages} {056037} (\bibinfo {year} {2024})}\BibitemShut {NoStop}%
\bibitem [{\citenamefont {Lees}\ \emph {et~al.}(2017)\citenamefont {Lees} \emph {et~al.}}]{BaBar:2017zmc}%
  \BibitemOpen
  \bibfield  {author} {\bibinfo {author} {\bibfnamefont {J.~P.}\ \bibnamefont {Lees}} \emph {et~al.} (\bibinfo {collaboration} {BaBar}),\ }\bibfield  {title} {\bibinfo {title} {{Measurement of the ${e}^{+}{e}^{{-}}{\rightarrow}{{\pi}}^{+}{{\pi}}^{{-}}{{\pi}}^{0}{{\pi}}^{0}$ cross section using initial-state radiation at BABAR}},\ }\href {https://doi.org/10.1103/PhysRevD.96.092009} {\bibfield  {journal} {\bibinfo  {journal} {Phys. Rev. D}\ }\textbf {\bibinfo {volume} {96}},\ \bibinfo {pages} {092009} (\bibinfo {year} {2017})}\BibitemShut {NoStop}%
\bibitem [{\citenamefont {Abdallah}\ and\ \citenamefont {Others}(2010)}]{DELPHICollab}%
  \BibitemOpen
  \bibfield  {author} {\bibinfo {author} {\bibfnamefont {J.}~\bibnamefont {Abdallah}}\ and\ \bibinfo {author} {\bibnamefont {Others}} (\bibinfo {collaboration} {The DELPHI Collaboration}),\ }\bibfield  {title} {\bibinfo {title} {Measurements of cp-conserving trilinear gauge boson couplings $wwv$ ($v\equiv \gamma, z$) in $e^{+}e^{-}$ collisions at lep2},\ }\href {https://doi.org/10.1140/epjc/s10052-010-1254-1} {\bibfield  {journal} {\bibinfo  {journal} {The European Physical Journal C}\ }\textbf {\bibinfo {volume} {66}},\ \bibinfo {pages} {35} (\bibinfo {year} {2010})}\BibitemShut {NoStop}%
\bibitem [{\citenamefont {Mele}(2004)}]{MELE2004255}%
  \BibitemOpen
  \bibfield  {author} {\bibinfo {author} {\bibfnamefont {S.}~\bibnamefont {Mele}},\ }\bibfield  {title} {\bibinfo {title} {Physics of w bosons at lep},\ }\href {https://doi.org/https://doi.org/10.1016/j.physrep.2004.08.017} {\bibfield  {journal} {\bibinfo  {journal} {Physics Reports}\ }\textbf {\bibinfo {volume} {403-404}},\ \bibinfo {pages} {255} (\bibinfo {year} {2004})}\BibitemShut {NoStop}%
\bibitem [{\citenamefont {Badalian}\ and\ \citenamefont {Simonov}(2013)}]{Badalian:2012ft}%
  \BibitemOpen
  \bibfield  {author} {\bibinfo {author} {\bibfnamefont {A.~M.}\ \bibnamefont {Badalian}}\ and\ \bibinfo {author} {\bibfnamefont {Y.~A.}\ \bibnamefont {Simonov}},\ }\bibfield  {title} {\bibinfo {title} {{Magnetic moments of mesons}},\ }\href {https://doi.org/10.1103/PhysRevD.87.074012} {\bibfield  {journal} {\bibinfo  {journal} {Phys. Rev. D}\ }\textbf {\bibinfo {volume} {87}},\ \bibinfo {pages} {074012} (\bibinfo {year} {2013})}\BibitemShut {NoStop}%
\bibitem [{\citenamefont {Lee}\ \emph {et~al.}(2008)\citenamefont {Lee}, \citenamefont {Moerschbacher},\ and\ \citenamefont {Wilcox}}]{Lee:2008qf}%
  \BibitemOpen
  \bibfield  {author} {\bibinfo {author} {\bibfnamefont {F.~X.}\ \bibnamefont {Lee}}, \bibinfo {author} {\bibfnamefont {S.}~\bibnamefont {Moerschbacher}},\ and\ \bibinfo {author} {\bibfnamefont {W.}~\bibnamefont {Wilcox}},\ }\bibfield  {title} {\bibinfo {title} {{Magnetic moments of vector, axial, and tensor mesons in lattice QCD}},\ }\href {https://doi.org/10.1103/PhysRevD.78.094502} {\bibfield  {journal} {\bibinfo  {journal} {Phys. Rev. D}\ }\textbf {\bibinfo {volume} {78}},\ \bibinfo {pages} {094502} (\bibinfo {year} {2008})}\BibitemShut {NoStop}%
\bibitem [{\citenamefont {Garcia~Gudino}\ and\ \citenamefont {Toledo~Sanchez}(2010)}]{GarciaGudino:2010sd}%
  \BibitemOpen
  \bibfield  {author} {\bibinfo {author} {\bibfnamefont {D.}~\bibnamefont {Garcia~Gudino}}\ and\ \bibinfo {author} {\bibfnamefont {G.}~\bibnamefont {Toledo~Sanchez}},\ }\bibfield  {title} {\bibinfo {title} {{Finite width induced modification to the electromagnetic form factors of spin-1 particles}},\ }\href {https://doi.org/10.1103/PhysRevD.81.073006} {\bibfield  {journal} {\bibinfo  {journal} {Phys. Rev. D}\ }\textbf {\bibinfo {volume} {81}},\ \bibinfo {pages} {073006} (\bibinfo {year} {2010})}\BibitemShut {NoStop}%
\bibitem [{\citenamefont {Lees}\ \emph {et~al.}(2012)\citenamefont {Lees} \emph {et~al.}}]{BaBar:2011btv}%
  \BibitemOpen
  \bibfield  {author} {\bibinfo {author} {\bibfnamefont {J.~P.}\ \bibnamefont {Lees}} \emph {et~al.} (\bibinfo {collaboration} {BaBar}),\ }\bibfield  {title} {\bibinfo {title} {{Cross Sections for the Reactions e+e- --{\ensuremath{>}} K+ K- pi+pi-, K+ K- pi0pi0, and K+ K- K+ K- Measured Using Initial-State Radiation Events}},\ }\href {https://doi.org/10.1103/PhysRevD.86.012008} {\bibfield  {journal} {\bibinfo  {journal} {Phys. Rev. D}\ }\textbf {\bibinfo {volume} {86}},\ \bibinfo {pages} {012008} (\bibinfo {year} {2012})}\BibitemShut {NoStop}%
\bibitem [{\citenamefont {Kroll}\ \emph {et~al.}(1967)\citenamefont {Kroll}, \citenamefont {Lee},\ and\ \citenamefont {Zumino}}]{Kroll:1967it}%
  \BibitemOpen
  \bibfield  {author} {\bibinfo {author} {\bibfnamefont {N.~M.}\ \bibnamefont {Kroll}}, \bibinfo {author} {\bibfnamefont {T.~D.}\ \bibnamefont {Lee}},\ and\ \bibinfo {author} {\bibfnamefont {B.}~\bibnamefont {Zumino}},\ }\bibfield  {title} {\bibinfo {title} {{Neutral Vector Mesons and the Hadronic Electromagnetic Current}},\ }\href {https://doi.org/10.1103/PhysRev.157.1376} {\bibfield  {journal} {\bibinfo  {journal} {Phys. Rev.}\ }\textbf {\bibinfo {volume} {157}},\ \bibinfo {pages} {1376} (\bibinfo {year} {1967})}\BibitemShut {NoStop}%
\bibitem [{\citenamefont {Bando}\ \emph {et~al.}(1985)\citenamefont {Bando}, \citenamefont {Kugo}, \citenamefont {Uehara}, \citenamefont {Yamawaki},\ and\ \citenamefont {Yanagida}}]{Bando:1984ej}%
  \BibitemOpen
  \bibfield  {author} {\bibinfo {author} {\bibfnamefont {M.}~\bibnamefont {Bando}}, \bibinfo {author} {\bibfnamefont {T.}~\bibnamefont {Kugo}}, \bibinfo {author} {\bibfnamefont {S.}~\bibnamefont {Uehara}}, \bibinfo {author} {\bibfnamefont {K.}~\bibnamefont {Yamawaki}},\ and\ \bibinfo {author} {\bibfnamefont {T.}~\bibnamefont {Yanagida}},\ }\bibfield  {title} {\bibinfo {title} {{Is rho Meson a Dynamical Gauge Boson of Hidden Local Symmetry?}},\ }\href {https://doi.org/10.1103/PhysRevLett.54.1215} {\bibfield  {journal} {\bibinfo  {journal} {Phys. Rev. Lett.}\ }\textbf {\bibinfo {volume} {54}},\ \bibinfo {pages} {1215} (\bibinfo {year} {1985})}\BibitemShut {NoStop}%
\bibitem [{\citenamefont {Fujiwara}\ \emph {et~al.}(1985)\citenamefont {Fujiwara}, \citenamefont {Kugo}, \citenamefont {Terao}, \citenamefont {Uehara},\ and\ \citenamefont {Yamawaki}}]{Fujiwara:1984mp}%
  \BibitemOpen
  \bibfield  {author} {\bibinfo {author} {\bibfnamefont {T.}~\bibnamefont {Fujiwara}}, \bibinfo {author} {\bibfnamefont {T.}~\bibnamefont {Kugo}}, \bibinfo {author} {\bibfnamefont {H.}~\bibnamefont {Terao}}, \bibinfo {author} {\bibfnamefont {S.}~\bibnamefont {Uehara}},\ and\ \bibinfo {author} {\bibfnamefont {K.}~\bibnamefont {Yamawaki}},\ }\bibfield  {title} {\bibinfo {title} {{Nonabelian Anomaly and Vector Mesons as Dynamical Gauge Bosons of Hidden Local Symmetries}},\ }\href {https://doi.org/10.1143/PTP.73.926} {\bibfield  {journal} {\bibinfo  {journal} {Prog. Theor. Phys.}\ }\textbf {\bibinfo {volume} {73}},\ \bibinfo {pages} {926} (\bibinfo {year} {1985})}\BibitemShut {NoStop}%
\bibitem [{\citenamefont {Meißner}(1988)}]{Meissner:1987ge}%
  \BibitemOpen
  \bibfield  {author} {\bibinfo {author} {\bibfnamefont {U.-G.}\ \bibnamefont {Meißner}},\ }\bibfield  {title} {\bibinfo {title} {{Low-Energy Hadron Physics from Effective Chiral Lagrangians with Vector Mesons}},\ }\href {https://doi.org/10.1016/0370-1573(88)90090-7} {\bibfield  {journal} {\bibinfo  {journal} {Phys. Rept.}\ }\textbf {\bibinfo {volume} {161}},\ \bibinfo {pages} {213} (\bibinfo {year} {1988})}\BibitemShut {NoStop}%
\bibitem [{\citenamefont {Navas}\ \emph {et~al.}(2024)\citenamefont {Navas} \emph {et~al.}}]{pdg}%
  \BibitemOpen
  \bibfield  {author} {\bibinfo {author} {\bibfnamefont {S.}~\bibnamefont {Navas}} \emph {et~al.} (\bibinfo {collaboration} {Particle Data Group}),\ }\bibfield  {title} {\bibinfo {title} {{Review of particle physics}},\ }\href {https://doi.org/10.1103/PhysRevD.110.030001} {\bibfield  {journal} {\bibinfo  {journal} {Phys. Rev. D}\ }\textbf {\bibinfo {volume} {110}},\ \bibinfo {pages} {030001} (\bibinfo {year} {2024})}\BibitemShut {NoStop}%
\bibitem [{\citenamefont {Kim}\ and\ \citenamefont {Tsai}(1973)}]{Kim:1973ee}%
  \BibitemOpen
  \bibfield  {author} {\bibinfo {author} {\bibfnamefont {K.~J.}\ \bibnamefont {Kim}}\ and\ \bibinfo {author} {\bibfnamefont {Y.-S.}\ \bibnamefont {Tsai}},\ }\bibfield  {title} {\bibinfo {title} {{MAGNETIC DIPOLE AND ELECTRIC QUADRUPOLE MOMENTS OF W+- MESON}},\ }\href {https://doi.org/10.1103/PhysRevD.7.3710} {\bibfield  {journal} {\bibinfo  {journal} {Phys. Rev. D}\ }\textbf {\bibinfo {volume} {7}},\ \bibinfo {pages} {3710} (\bibinfo {year} {1973})}\BibitemShut {NoStop}%
\bibitem [{\citenamefont {Hagiwara}\ \emph {et~al.}(1987)\citenamefont {Hagiwara}, \citenamefont {Peccei}, \citenamefont {Zeppenfeld},\ and\ \citenamefont {Hikasa}}]{Hagiwara:1986vm}%
  \BibitemOpen
  \bibfield  {author} {\bibinfo {author} {\bibfnamefont {K.}~\bibnamefont {Hagiwara}}, \bibinfo {author} {\bibfnamefont {R.~D.}\ \bibnamefont {Peccei}}, \bibinfo {author} {\bibfnamefont {D.}~\bibnamefont {Zeppenfeld}},\ and\ \bibinfo {author} {\bibfnamefont {K.}~\bibnamefont {Hikasa}},\ }\bibfield  {title} {\bibinfo {title} {{Probing the Weak Boson Sector in e+ e- ---\ensuremath{>} W+ W-}},\ }\href {https://doi.org/10.1016/0550-3213(87)90685-7} {\bibfield  {journal} {\bibinfo  {journal} {Nucl. Phys. B}\ }\textbf {\bibinfo {volume} {282}},\ \bibinfo {pages} {253} (\bibinfo {year} {1987})}\BibitemShut {NoStop}%
\bibitem [{\citenamefont {Nieves}\ and\ \citenamefont {Pal}(1997)}]{Nieves:1996ff}%
  \BibitemOpen
  \bibfield  {author} {\bibinfo {author} {\bibfnamefont {J.~F.}\ \bibnamefont {Nieves}}\ and\ \bibinfo {author} {\bibfnamefont {P.~B.}\ \bibnamefont {Pal}},\ }\bibfield  {title} {\bibinfo {title} {{Electromagnetic properties of neutral and charged spin 1 particles}},\ }\href {https://doi.org/10.1103/PhysRevD.55.3118} {\bibfield  {journal} {\bibinfo  {journal} {Phys. Rev. D}\ }\textbf {\bibinfo {volume} {55}},\ \bibinfo {pages} {3118} (\bibinfo {year} {1997})}\BibitemShut {NoStop}%
\bibitem [{\citenamefont {Gounaris}\ \emph {et~al.}(1996)\citenamefont {Gounaris} \emph {et~al.}}]{Gounaris:1996rz}%
  \BibitemOpen
  \bibfield  {author} {\bibinfo {author} {\bibfnamefont {G.}~\bibnamefont {Gounaris}} \emph {et~al.},\ }\bibfield  {title} {\bibinfo {title} {Triple gauge boson couplings},\ }in\ \href {https://doi.org/10.48550/arXiv.hep-ph/9601233} {\emph {\bibinfo {booktitle} {AGS / RHIC Users Annual Meeting}}}\ (\bibinfo {year} {1996})\ \Eprint {https://arxiv.org/abs/hep-ph/9601233} {arXiv:hep-ph/9601233} \BibitemShut {NoStop}%
\bibitem [{\citenamefont {Baur}\ and\ \citenamefont {Zeppenfeld}(1995)}]{Baur:1995aa}%
  \BibitemOpen
  \bibfield  {author} {\bibinfo {author} {\bibfnamefont {U.}~\bibnamefont {Baur}}\ and\ \bibinfo {author} {\bibfnamefont {D.}~\bibnamefont {Zeppenfeld}},\ }\bibfield  {title} {\bibinfo {title} {{Finite width effects and gauge invariance in radiative $W$ productions and decay}},\ }\href {https://doi.org/10.1103/PhysRevLett.75.1002} {\bibfield  {journal} {\bibinfo  {journal} {Phys. Rev. Lett.}\ }\textbf {\bibinfo {volume} {75}},\ \bibinfo {pages} {1002} (\bibinfo {year} {1995})}\BibitemShut {NoStop}%
\bibitem [{\citenamefont {Argyres}\ \emph {et~al.}(1995)\citenamefont {Argyres}, \citenamefont {Beenakker}, \citenamefont {van Oldenborgh}, \citenamefont {Denner}, \citenamefont {Dittmaier}, \citenamefont {Hoogland}, \citenamefont {Kleiss}, \citenamefont {Papadopoulos},\ and\ \citenamefont {Passarino}}]{Argyres:1995ym}%
  \BibitemOpen
  \bibfield  {author} {\bibinfo {author} {\bibfnamefont {E.~N.}\ \bibnamefont {Argyres}}, \bibinfo {author} {\bibfnamefont {W.}~\bibnamefont {Beenakker}}, \bibinfo {author} {\bibfnamefont {G.~J.}\ \bibnamefont {van Oldenborgh}}, \bibinfo {author} {\bibfnamefont {A.}~\bibnamefont {Denner}}, \bibinfo {author} {\bibfnamefont {S.}~\bibnamefont {Dittmaier}}, \bibinfo {author} {\bibfnamefont {J.}~\bibnamefont {Hoogland}}, \bibinfo {author} {\bibfnamefont {R.}~\bibnamefont {Kleiss}}, \bibinfo {author} {\bibfnamefont {C.~G.}\ \bibnamefont {Papadopoulos}},\ and\ \bibinfo {author} {\bibfnamefont {G.}~\bibnamefont {Passarino}},\ }\bibfield  {title} {\bibinfo {title} {{Stable calculations for unstable particles: Restoring gauge invariance}},\ }\href {https://doi.org/10.1016/0370-2693(95)01002-8} {\bibfield  {journal} {\bibinfo  {journal} {Phys. Lett. B}\ }\textbf {\bibinfo {volume} {358}},\ \bibinfo {pages} {339} (\bibinfo {year} {1995})}\BibitemShut {NoStop}%
\bibitem [{\citenamefont {Beuthe}\ \emph {et~al.}(1997)\citenamefont {Beuthe}, \citenamefont {Gonzalez~Felipe}, \citenamefont {Lopez~Castro},\ and\ \citenamefont {Pestieau}}]{Beuthe:1996fe}%
  \BibitemOpen
  \bibfield  {author} {\bibinfo {author} {\bibfnamefont {M.}~\bibnamefont {Beuthe}}, \bibinfo {author} {\bibfnamefont {R.}~\bibnamefont {Gonzalez~Felipe}}, \bibinfo {author} {\bibfnamefont {G.}~\bibnamefont {Lopez~Castro}},\ and\ \bibinfo {author} {\bibfnamefont {J.}~\bibnamefont {Pestieau}},\ }\bibfield  {title} {\bibinfo {title} {{Behavior of the absorptive part of the W+- electromagnetic vertex}},\ }\href {https://doi.org/10.1016/S0550-3213(97)00263-0} {\bibfield  {journal} {\bibinfo  {journal} {Nucl. Phys. B}\ }\textbf {\bibinfo {volume} {498}},\ \bibinfo {pages} {55} (\bibinfo {year} {1997})}\BibitemShut {NoStop}%
\bibitem [{\citenamefont {Lopez~Castro}\ and\ \citenamefont {Toledo~Sanchez}(2000)}]{LopezCastro:1999xg}%
  \BibitemOpen
  \bibfield  {author} {\bibinfo {author} {\bibfnamefont {G.}~\bibnamefont {Lopez~Castro}}\ and\ \bibinfo {author} {\bibfnamefont {G.}~\bibnamefont {Toledo~Sanchez}},\ }\bibfield  {title} {\bibinfo {title} {{Gauge invariance and finite width effects in radiative two pion tau lepton decay}},\ }\href {https://doi.org/10.1103/PhysRevD.61.033007} {\bibfield  {journal} {\bibinfo  {journal} {Phys. Rev. D}\ }\textbf {\bibinfo {volume} {61}},\ \bibinfo {pages} {033007} (\bibinfo {year} {2000})}\BibitemShut {NoStop}%
\bibitem [{\citenamefont {Kumar}(1969)}]{Kumar:1969jjy}%
  \BibitemOpen
  \bibfield  {author} {\bibinfo {author} {\bibfnamefont {R.}~\bibnamefont {Kumar}},\ }\bibfield  {title} {\bibinfo {title} {{Covariant phase-space calculations of n-body decay and production processes}},\ }\href {https://doi.org/10.1103/PhysRev.185.1865} {\bibfield  {journal} {\bibinfo  {journal} {Phys. Rev.}\ }\textbf {\bibinfo {volume} {185}},\ \bibinfo {pages} {1865} (\bibinfo {year} {1969})}\BibitemShut {NoStop}%
\bibitem [{\citenamefont {{Peter Lepage}}(1978)}]{PETER1978192}%
  \BibitemOpen
  \bibfield  {author} {\bibinfo {author} {\bibfnamefont {G.}~\bibnamefont {{Peter Lepage}}},\ }\bibfield  {title} {\bibinfo {title} {A new algorithm for adaptive multidimensional integration},\ }\href {https://doi.org/https://doi.org/10.1016/0021-9991(78)90004-9} {\bibfield  {journal} {\bibinfo  {journal} {Journal of Computational Physics}\ }\textbf {\bibinfo {volume} {27}},\ \bibinfo {pages} {192} (\bibinfo {year} {1978})}\BibitemShut {NoStop}%
\bibitem [{\citenamefont {\'Avalos}\ \emph {et~al.}(2023)\citenamefont {\'Avalos}, \citenamefont {Rojas}, \citenamefont {S\'anchez},\ and\ \citenamefont {Toledo}}]{Avalos:2022txl}%
  \BibitemOpen
  \bibfield  {author} {\bibinfo {author} {\bibfnamefont {G.}~\bibnamefont {\'Avalos}}, \bibinfo {author} {\bibfnamefont {A.}~\bibnamefont {Rojas}}, \bibinfo {author} {\bibfnamefont {M.}~\bibnamefont {S\'anchez}},\ and\ \bibinfo {author} {\bibfnamefont {G.}~\bibnamefont {Toledo}},\ }\bibfield  {title} {\bibinfo {title} {{Role of the \ensuremath{\rho}(1450) in low-energy observables from an analysis in the meson dominance approach}},\ }\href {https://doi.org/10.1103/PhysRevD.107.056006} {\bibfield  {journal} {\bibinfo  {journal} {Phys. Rev. D}\ }\textbf {\bibinfo {volume} {107}},\ \bibinfo {pages} {056006} (\bibinfo {year} {2023})}\BibitemShut {NoStop}%
\end{thebibliography}%

\end{document}